\documentclass[%
 reprint,
superscriptaddress,
nofootinbib,
 amsmath,amssymb,
 aps,
]{revtex4-1}

\usepackage{graphicx}% Include figure files
\usepackage{dcolumn}% Align table columns on decimal point
\usepackage{bm}% bold math
\usepackage{subfigure}

\begin{document}

\preprint{APS/123-QED}

\title{Hidden Markov model tracking of continuous gravitational waves from a neutron star \\ with wandering spin}% Force line breaks with \\

\author{S. Suvorova}
\affiliation{School of Electrical and Computer Engineering, RMIT University, Melbourne, Victoria 3000, Australia}
\affiliation{School of Physics, University of Melbourne, Parkville, Victoria 3010, Australia}
\author{L. Sun}
\email{lings2@student.unimelb.edu.au}
\author{A. Melatos}
\email{amelatos@unimelb.edu.au}
\affiliation{School of Physics, University of Melbourne, Parkville, Victoria 3010, Australia}%Lines break automatically or can be forced with \\
\author{W. Moran}
\affiliation{School of Electrical and Computer Engineering, RMIT University, Melbourne, Victoria 3000, Australia}
\author{R. J. Evans}
\affiliation{Department of Electrical and Electronic Engineering, University of Melbourne, Parkville, Victoria 3010, Australia}

\date{\today}% It is always \today, today,

\begin{abstract}
Gravitational wave searches for continuous-wave signals from neutron stars are especially challenging when the star's spin frequency is unknown a priori from electromagnetic observations and wanders stochastically under the action of internal (e.g. superfluid or magnetospheric) or external (e.g. accretion) torques. It is shown that frequency tracking by hidden Markov model (HMM) methods can be combined with existing maximum likelihood coherent matched filters like the $\mathcal{F}$-statistic to surmount some of the challenges raised by spin wandering. Specifically it is found that, for an isolated, biaxial rotor whose spin frequency walks randomly, HMM tracking of the $\mathcal{F}$-statistic output from coherent segments with duration $T_\text{drift}=10$\,d over a total observation time of $T_\text{obs}=1$\,yr can detect signals with wave strains $h_0 > 2 \times 10^{-26}$ at a noise level characteristic of the Advanced Laser Interferometer Gravitational Wave Observatory (Advanced LIGO). For a biaxial rotor with randomly walking spin in a binary orbit, whose orbital period and semi-major axis are known approximately from electromagnetic observations, HMM tracking of the Bessel-weighted $\mathcal{F}$-statistic output can detect signals with $h_0 > 8 \times 10^{-26}$. An efficient, recursive, HMM solver based on the Viterbi algorithm is demonstrated, which requires $\sim 10^3$ CPU-hours for a typical, broadband (0.5-kHz) search for the low-mass X-ray binary Scorpius X-1, including generation of the relevant $\mathcal{F}$-statistic input. In a ``realistic" observational scenario, Viterbi tracking successfully detects 41 out of 50 synthetic signals without spin wandering in Stage I of the Scorpius X-1 Mock Data Challenge convened by the LIGO Scientific Collaboration down to a wave strain of $h_0=1.1 \times 10^{-25}$, recovering the frequency with a root-mean-square accuracy of $\leq4.3 \times10^{-3}$\,Hz.

\begin{description}
\item[PACS numbers]
95.85.Sz, 97.60.Jd
\end{description}
\end{abstract}

\pacs{Valid PACS appear here}% PACS, the Physics and Astronomy
\maketitle

%\tableofcontents

\section{\label{sec:level1}Introduction}

Continuous-wave gravitational radiation from isolated and accreting neutron stars is a key target of long-baseline interferometers like the Laser Interferometer Gravitational Wave Observatory (LIGO) and Virgo detector \cite{Riles2013}. Theory predicts that the signal is quasi-monochromatic. Emission occurs at simple rational multiples of the star's spin frequency $f_\star$, for example $f_\star$ and $2f_\star$ for mass quadrupole radiation from thermoelastic and magnetic mountains \cite{Ushomirsky2000,Melatos2005}, $4f_\star/3$ for r-modes \cite{Owen1998,Bondarescu2009}, and $f_\star$ for current quadrupole radiation from nonaxisymmetric flows in the neutron superfluid pinned to the stellar crust \cite{Melatos2015}. If the source exhibits electromagnetic pulsations, so that an ephemeris can be derived from absolute pulse numbering, i.e. $f_\star(t)$ is known as a function of time $t$, it is customary to search for a signal using coherent matched filters like the maximum likelihood $\mathcal{F}$-statistic \cite{Jaranowski1998}. If an ephemeris is unavailable, coherent searches over multiple $f_\star (t)$ templates indexed by the Taylor coefficients $f^{(k)}_\star(0)=(d^kf_\star/dt^k)_{t=0}$ become expensive computationally \cite{Jones2015}, and semi-coherent methods are often preferred. Cross-correlation \cite{Dhurandhar2008,Chung2011}, StackSlide \cite{StackSlide2005}, the Hough transform \cite{HoughKrishnan2004,HoughSearchAasi2013,Aasi2016}, PowerFlux \cite{PowerFlux2005a,PowerFlux2005b,PowerFlux2011,PowerFluxSearch2009,PowerFluxSearch2012,Aasi2016-PowerFlux} and TwoSpect \cite{TwoSpectGoetz2011,TwoSpectSearchAasi2014} are all examples of semi-coherent algorithms implemented by the LIGO Scientific Collaboration and applied to data from Science Runs 5 or 6.

Search methods that scan templates without guidance from a measured ephemeris are compromised if $f_\star(t)$ wanders randomly. Radio and X-ray timing of pulsating neutron stars reveal that spin wandering is a widespread phenomenon. In isolated objects, it manifests itself as timing noise \cite{Hobbs2010,Shannon2010,Ashton2015}, exhibits a red Fourier spectrum with an auto-correlation time-scale of days to years \cite{Cordes1980,Price2012}, and has been attributed variously to magnetospheric changes \cite{Lyne2010}, superfluid dynamics in the stellar interior \cite{Alpar1986,Jones1990,Price2012,Melatos2014}, spin microjumps \cite{Cordes1985,Janssen2006}, and fluctuations in the spin-down torque \cite{Cheng1987,Cheng1987a,Urama2006}. In accreting objects, spin wandering results from fluctuations in the magnetized accretion torque \cite{DeKool1993,Baykal1993,Bildsten1997}, due to transient accretion disk formation \cite{Taam1988,Baykal1991} or  disk-magnetospheric instabilities and reconnection events \cite{Romanova2004}. Again the auto-correlation time-scale is of the order of days \cite{Baykal1997}, and fluctuations in $f_\star(t)$ are accompanied by fluctuations in the X-ray flux \cite{Bildsten1997}. Even when the ephemeris is measured electromagnetically, the gravitational-wave-emitting quadrupole may not be phase locked to the stellar crust, and wandering must still be accommodated \cite{Crab2008}.

Hidden Markov model (HMM) methods offer one powerful strategy for detecting and tracking a wandering frequency \cite{Quinn2001}. The essential idea is to model $f_\star(t)$ probabilistically as a Markov chain of transitions between unobservable (``hidden") frequency states and relate the hidden states to the observed data via a detection statistic. HMM frequency tracking enjoys a long track record of success in engineering applications ranging from radar and sonar analysis \cite{Paris2003} to mobile telephony \cite{White2002,Williams2002}. It has been refined substantially since its introduction by \citet{Streit1990} to include information about amplitude and phase \cite{Barrett1993} and embrace simultaneous tracking of multiple targets and frequencies \cite{Xie1991,Xie1993}. It delivers accurate estimation, when the signal-to-noise ratio (SNR) is low but the sample size is large \cite{Quinn2001}, the situation normally confronting gravitational wave data analysis targeting continuous-wave sources.

In this paper, we implement and test a specific HMM scheme based on the classic Viterbi algorithm \cite{Viterbi1967,Quinn2001}. The scheme is efficient and practical: its computational demands are modest, and it co-opts existing technology for LIGO continuous-wave searches based on the $\mathcal{F}$-statistic. The paper is organized as follows. In Section II, we formulate the search problem in HMM language and describe how to solve it with the Viterbi algorithm. In Section III, we apply Viterbi tracking to an isolated neutron star with $f_\star(t)$ walking randomly, using the $\mathcal{F}$-statistic to relate the hidden and observable states. We quantify the performance of the tracker as a function of signal strength using synthetic data. In Section IV we apply Viterbi tracking to a neutron star in a binary, again with $f_\star(t)$ wandering randomly, replacing the $\mathcal{F}$-statistic with a Bessel-weighted variant, and quantify its performance using synthetic data. Finally, to illustrate how the tracker performs in a ``realistic" scenario, we apply it to the data set prepared for Stage I of the Scorpius X-1 (Sco X-1) Mock Data Challenge in Section V \cite{Messenger2015}.

\section{Frequency Tracking}

HMM principles can be applied to gravitational wave frequency tracking in various ways, depending on the data format, the detection statistic, and any known constraints on the frequency evolution. 

In this paper, we specialize to continuous-wave searches, where the raw data are packaged in 30-min short Fourier transforms (SFTs), during which $f_\star(t)$ remains localized within one Fourier bin. The SFTs are fed into a frequency-domain estimator $G(f)$, like the $\mathcal{F}$-statistic (isolated target) or Bessel-weighted $\mathcal{F}$-statistic (binary target). It is safe to assume that $f_\star(t)$ remains localized within one estimator frequency bin over a short enough time interval, even when $f_\star(t)$ walks randomly. Let $T_\text{obs}$ be the total observation time, and let $T_\text{SFT}=30$\,min be the length of each SFT. For any particular astrophysical source, there exists an intermediate time-scale $T_\text{drift}$ ($T_\text{SFT} < T_\text{drift} < T_\text{obs}$), over which $f_\star(t)$ wanders by at most one estimator bin. For timing noise in isolated pulsars, radio timing experiments imply $T_\text{drift} \gtrsim \text{weeks}$ \cite{Hobbs2010}. For accretion noise in low-mass X-ray binaries, such as Sco X-1, one can estimate $T_\text{drift} \approx 10$\,days theoretically \cite{Sammut2014,SCO-X1-2015,Messenger2015}. Our general strategy is to compute $G(f)$ for blocks of data of length $T_\text{drift}$ then use the Viterbi algorithm to track peaks in $G(f)$ over the full observation interval $0 \leq t \leq T_\text{obs}$, effectively summing the estimator output semi-coherently by tracking $f_\star(t)$.\footnote{$T_\text{drift}$ can be as short as $T_\text{SFT}$, the minimum time over which it makes sense to calculate $G(f)$, without harming the performance of the tracking algorithm. As the algorithm is semi-coherent, the strain sensitivity scales approximately as $(T_\text{obs}/T_\text{drift})^{1/4}$ \cite{Riles2013}.} An important practical advantage of this approach is that it leverages the existing, efficient, thoroughly tested $\mathcal{F}$-statistic software infrastructure within the LIGO Algorithms Library (LAL), which is used extensively by the continuous-wave data analysis community \cite{F-stat2011}. 

In this section we formulate the problem as an HMM (Section \ref{sec:hidden_markov}--\ref{sec:setup_markov}) and describe the recursive Viterbi algorithm for solving the problem (Section \ref{sec:viterbi}). We defer to future work the treatment of glitches, i.e. random, impulsive, spin-up events, where $f_\star (t)$ jumps over many estimator bins instantaneously \cite{Melatos2008,Espinoza2011}.

\subsection{HMM formulation}
\label{sec:hidden_markov}

An HMM is a probabilistic finite state automaton defined by a hidden (unobservable) state variable $q(t)$, which takes one of a finite set of values $\{q_1, \cdots, q_{N_Q}\}$ at time $t$, and an observable state variable $o(t)$, which takes one of the values $\{o_1, \cdots, o_{N_o}\}$. The automaton jumps between states at discrete times $\{t_0, \cdots, t_{N_T}\}$. The jump probability from time $t_n$ to time $t_{n+1}$ depends only on the hidden state $q(t_n)$ at time $t_n$ --- the Markovian assumption --- and is described by the transition probability matrix
\begin{equation}
\label{eqn:prob_matrix}
A_{q_j q_i} = \Pr [q(t_{n+1})=q_j|q(t_n)=q_i].
\end{equation}
The likelihood that the system is observed in state $o(t_n)$ at time $t_n$ is described by the emission probability matrix
\begin{equation}
\label{eqn:likelihood}
L_{o_j q_i} = \Pr [o(t_n)=o_j|q(t_n)=q_i].
\end{equation}
The model is completed by specifying the probability that the system occupies each hidden state initially, described by the prior vector
\begin{equation}
\Pi_{q_i} = \Pr [q(t_0)=q_i].
\end{equation}

Suppose that we observe the system transitioning through the sequence of observable states $O=\{o(t_0), \cdots, o(t_{N_T})\}$. In general there exist $N_Q^{N_T + 1}$ possible paths $Q=\{q(t_0), \cdots, q(t_{N_T})\}$ through the hidden states which are consistent with the observed sequence $O$. For a Markov process, the probability that the hidden path $Q$ gives rise to the observed sequence $O$ equals
\begin{equation}
\label{eqn:prob}
\begin{split}
P(Q|O) = & L_{o(t_{N_T})q(t_{N_T})} A_{q(t_{N_T})q(t_{N_T-1})} \cdots L_{o(t_1)q(t_1)} \\ 
& \times A_{q(t_1)q(t_0)} \Pi_{q(t_0)}.
\end{split}
\end{equation}
The most probable path $Q^*$ is the one that maximizes $P(Q|O)$, i.e., 
\begin{equation}
\label{eqn:best_path}
Q^*(O) = \arg\max P(Q|O),
\end{equation}
where $\arg\max (\ldots)$ returns the argument that maximizes the function $(\ldots)$. The Viterbi algorithm, presented in Section \ref{sec:viterbi}, provides a recursive, computationally efficient route to computing $Q^*(O)$ from (\ref{eqn:prob_matrix})--(\ref{eqn:best_path}). For computational reasons, we actually evaluate $\log P(Q|O)$, whereupon (\ref{eqn:prob}) becomes a sum of $\log$ likelihoods.

\subsection{Frequency drift time-scale and jump probabilities}
\label{sec:jump_prob}
In our first application to isolated neutron stars (Section \ref{sec:isolated_ns}), the hidden state variable is one-dimensional and equals the star's spin frequency at time $t$, i.e. $q(t)=f_\star(t)$. Its allowed, discretized values correspond one-to-one to the frequency bins in the output of the frequency-domain estimator $G(f_\star)$ computed over an interval of length $T_\text{drift}$, indexed by their central (say) frequencies ${f_\star}_i$, i.e. $q_i={f_\star}_i$.\footnote{If we track amplitude $h(t)$ and frequency $f_\star(t)$ jointly (outside the scope of this paper), then the state space enlarges appropriately, with $q(t)=[f_\star(t),h(t)].$} If the total search band covers the frequency range $f_{\star, \text{min}} \leq f_\star(t) \leq f_{\star, \text{max}} = f_{\star, \text{min}} + B$, and if the bin width of $G(f_\star)$ is $\Delta f_\text{drift} = 1/(2 T_\text{drift})$, then the number of hidden states is given by $N_Q = B/\Delta f_\text{drift} $. 

Radio and X-ray timing observations demonstrate that neutron star spin wandering is a continuous stochastic process in the absence of glitches \cite{Bildsten1997,Hobbs2010,Shannon2010}. In what follows we approximate it by an unbiased random walk or Wiener process. Choosing $T_\text{drift}$ to satisfy
\begin{equation}
\left|\int_t^{t+T_\text{drift}}dt' \dot{f_\star}(t')\right| < \Delta f_\text{drift}
\end{equation}
guarantees that the transition probability matrix takes a simple tridiagonal form, with
\begin{equation}
\label{eqn:trans_matrix}
A_{q_{i+1} q_i} = A_{q_i q_i} = A_{q_{i-1} q_i} = \frac{1}{3}
\end{equation}
and all other entries zero. In other words, at each time step, $f_\star(t)$ jumps at most one frequency bin up or down or stays in the same bin with equal probability $1/3$. Studies show that the performance of a HMM tracking scheme is insensitive to the exact form of the probabilities $A_{q_j q_i}$, as long as they capture broadly the behaviour of the underlying jump process \cite{Quinn2001}. Equation (\ref{eqn:trans_matrix}) can be generalized to accommodate discontinuous glitches, but doing so lies outside the scope of this paper. \footnote{In one simple glitch model, let $\sigma = T_\text{drift}/T_\text{glitch}$ be the probability of a glitch occurring in an interval $T_\text{drift}$, where $T_\text{glitch}$ is the mean glitch waiting time, and let $\eta$ be the maximum fractional glitch size. Then we replace $1/3$ in equation (\ref{eqn:trans_matrix}) with $A_{q_{i-1} q_i}=(1-\sigma)/3, A_{q_{i+1} q_i} = A_{q_i q_i}=(1-\sigma)/3+\sigma (\eta q_i/\Delta f_\text{drift})^{-1}$ and also have $A_{q_j q_i}=\sigma (\eta q_i/\Delta f_\text{drift})^{-1}$ for $1 < q_j-q_i \leq \eta q_i$.}

In our second application to binary neutron stars (Sections \ref{sec:binary} and \ref{sec:MDC}), the hidden state variable is two-dimensional: we search over not only $f_\star(t)$ but also the projected semimajor axis $a_0 = a \sin \iota$, where $a$ is the semimajor axis of the orbit, and $\iota$ is the angle of inclination. In practice we divide the one-standard-deviation error bar on $a_0$ from electromagnetic measurements into $N_{a_0}$ bins, whereupon the total number of hidden states equals $N_Q=N_{a_0} B /\Delta f_\text{drift}$. It is known astrophysically that $a_0$ does not change significantly over a typical observation ($T_\text{obs} \sim 1$\,yr), so the transition matrix is still given by equation (\ref{eqn:trans_matrix}), with the subscript $i$ indexing the frequency bin ${f_\star}_i$ but not $a_0$. In other words, there is no dependence on $a_0$ in the two-dimensional version of equation (\ref{eqn:trans_matrix}). Therefore no separate subscript is needed to index $a_0$.

\subsection{Emission and prior probabilities}
\label{sec:setup_markov}
The observable state variable $o(t)$ comprises the data collected during the interval $t \leq t' \leq t+T_\text{drift}$. Formally it is the vector $[x(t'_0), \cdots, x(t'_{N_\text{drift}})]$, whose dimension $N_\text{drift}+1$ equals the interferometer sampling frequency ($\approx 16$\,kHz) times $T_\text{drift}$, where $x(t)=h(t)+n(t)$ is the output of the interferometer signal channel, $h(t)$ denotes the gravitational wave strain, and $n(t)$ denotes the noise. The emission probability matrix is then given by
\begin{eqnarray}
\label{eqn:emi_prob_matrix}
L_{o(t), q_i} &=& \Pr [o(t)|{f_\star}_i \leq f_\star(t) \leq {f_\star}_i+\Delta f_\text{drift}]\\ 
\label{eqn:matrix_propto}
&\propto& \exp[G({f_\star}_i)],
\end{eqnarray}
where (\ref{eqn:matrix_propto}) follows from (\ref{eqn:emi_prob_matrix}) by the definition of the frequency domain estimator.

Since we have no initial knowledge of $f_\star(t)$, as prior we chose uniform distribution, i.e 
\begin{equation}
\Pi_{q_i} = N_Q^{-1},
\end{equation}
for all $q_i$.
The maximum entropy of uniform distribution serves us to avoid any unwarranted assumptions about the signal.

In binary neutron star applications, $L_{o(t), q_i}$ depends on $a_0$ as well as $f_\star$; an explicit formula is given in Section \ref{sec:bessel}. The prior is the product of a uniform prior in $f_\star$ and a uniform or Gaussian prior in $a_0$, the latter centred on the electromagnetically measured value with a standard deviation equal to the measurement uncertainty.

\subsection{Viterbi algorithm}
\label{sec:viterbi}

Let ${Q^*}^{(k)}=[q^*(t_0), \cdots, q^*(t_k)]$ be the first $k+1$ steps in the most probable path maximizing (\ref{eqn:prob}) for the observation sequence $O^{(k)} = [o(t_0), \cdots, o(t_k)]$. For a Markov process, ${Q^*}^{(k)}$ satisfies the nesting property ${Q^*}^{(k-1)} \subset {Q^*}^{(k)}$, i.e. ${Q^*}^{(k-1)}$ makes up the first $k$ steps of ${Q^*}^{(k)}$. This is a special case of the Principle of Optimality \cite{Bellman1957}: if ${Q^*}^{(k)}$ is optimal, then all subpaths within ${Q^*}^{(k)}$ must be optimal too. In other words, given the optimal path ${Q^*}^{(k)}=[q^*(t_0),\cdots, q^*(t_i), \cdots, q^*(t_j), \cdots, q^*(t_k) ]$, one must have $\Pr \{[q'(t_i), \cdots, q'(t_j)] | [o(t_i), \cdots, o(t_j)]\} \leq \Pr \{[q^*(t_i), \cdots, q^*(t_j)] | [o(t_i), \cdots, o(t_j)]\}$ for all possible choices of $[q'(t_i), \cdots, q'(t_j)]$.

When applied to equation (\ref{eqn:prob}), the Principle of Optimality naturally defines a recursive algorithm, proposed by Viterbi \cite{Viterbi1967}, to find $Q^*$ by backtracking. At every forward step $k >0$ in the recursion, the Viterbi algorithm eliminates all but $N_Q$ possible state sequences. The retained sequences end in different states by construction; if more than one sequence ends in a given state, the sequence with maximum $P[Q^{(k)}|O^{(k)}]$ is retained. 

At time $t_k$, we save each of the $N_Q$ maximum probabilities in the vector $\bm{\delta} (t_k)$ with components
\begin{equation}
\label{eqn:delta_t_k}
\delta_{q_i}(t_k) = \mathop{\max} \limits_{q_j} \Pr[q(t_k) = q_i | q(t_{k-1}) = q_j; O^{(k)}]
\end{equation}
and we save the state at $t_{k-1}$ leading to each retained sequence in the vector $\bm {\Phi}(t_k)$ with components
\begin{equation}
\label{eqn:Phi_t_k}
\Phi_{q_i}(t_k) = \mathop{\arg \max} \limits_{q_j} \Pr[q(t_k) = q_i | q(t_{k-1}) = q_j; O^{(k)}],
\end{equation}
with 
\begin{equation}
\Pr [q(t_k) = q_i | q(t_{k-1}) = q_j; O^{(k)}] = L_{o(t_k)q_i} A_{q_i q_j}\delta_{q_j}(t_{k-1})
\end{equation}
in both (\ref{eqn:delta_t_k}) and (\ref{eqn:Phi_t_k}). After stepping $k$ forward through $1 \leq k \leq N_T$, we backtrack through the $\bm {\Phi} (t_k)$ vectors to find the optimal path.

In summary, therefore, the Viterbi algorithm comprises four stages.

$\emph{1. Initialization:}$
\begin{equation}
\delta_{q_i}(t_0) = L_{o(t_0)q_i} \Pi _{q_i},
\end{equation}
for $1 \leq i \leq N_Q$. Note that $\Phi_{q_i}(t_0)$ is never used.

$\emph{2. Recursion:}$
\begin{eqnarray}
\delta_{q_i}(t_k) &=& L_{o(t_k)q_i} \mathop{\max} \limits_{1 \leq j \leq N_Q} [A_{q_i q_j}\delta_{q_j}(t_{k-1})], \\
\Phi_{q_i}(t_k) &=& \mathop{\arg \max} \limits_{1 \leq j \leq N_Q} [A_{q_i q_j}\delta_{q_j}(t_{k-1})],
\end{eqnarray}
for $1 \leq i \leq N_Q$ and $1 \leq k \leq N_T$.

$\emph{3. Termination:}$
\begin{eqnarray}
\max P(Q|O) &=& \mathop{\max} \limits_{q_j} \delta_{q_j}(t_{N_T}) \\
q^*(t_{N_T}) &=& \mathop{\arg \max} \limits_{q_j} \delta_{q_j}(t_{N_T})
\end{eqnarray}
for $1 \leq j \leq N_Q$.

$\emph{4. Optimal path backtracking:}$
\begin{equation}
q^*(t_k) = \Phi_{q^*(t_{k+1})}(t_{k+1})
\end{equation}
for $0 \leq k \leq N_T -1$.

By pruning the tree of possible paths efficiently at each step, the Viterbi algorithm reduces the number of comparisons from $N_Q^{N_T+1}$ to $(N_T+1)N_Q^2$, which can be reduced further to $(N_T+1)N_Q \ln N_Q$ by binary maximization \cite{Quinn2001}.

\section{Isolated Neutron Star}
\label{sec:isolated_ns}
We first consider an isolated neutron star, whose spin frequency wanders randomly by at most plus or minus one $\mathcal{F}$-statistic frequency bin on the time-scale $T_\text{drift}$. We take $T_\text{obs}=370$\,d, $T_\text{drift}=10$\,d, and $\Delta f_\text{drift} = 5.787037 \times 10^{-7}$\,Hz for the sake of illustration and with an eye to later comparison with the tests in Section \ref{sec:binary} and \ref{sec:MDC} motivated by binaries like Sco X-1.

\subsection{Matched filter: $\mathcal{F}$-statistic}

Over time intervals that are short compared to $T_\text{drift}$, the optimal matched filter for a biaxial rotor with no orbital motion is the maximum-likelihood $\mathcal{F}$-statistic, which accounts for the rotation of the Earth and its orbit around the solar system barycentre (SSB).

The time-dependent data $x(t)$ collected at a single detector take the form
\begin{equation}
x(t)=\mathcal{A}^\mu h_\mu(t) + n(t),
\end{equation}
where $n(t)$ represents stationary, additive noise, $h_\mu(t)$ are the four linearly independent signal components
\begin{eqnarray}
\label{eqn:h1} h_1(t) &=& a(t) \cos \Phi(t), \\
h_2(t) &=& b(t) \cos \Phi(t), \\
h_3(t) &=& a(t) \sin \Phi(t),  \\
\label{eqn:h4} h_4(t) &=& b(t) \sin \Phi(t),
\end{eqnarray}
$a(t)$ and $b(t)$ are the antenna-pattern functions defined by Equations (12) and (13) in Ref. \cite{Jaranowski1998}, $\mathcal{A}^\mu$ is the amplitude associated with $h_\mu$, and $\Phi(t)$ is the signal phase at the detector. 

The $\mathcal{F}$-statistic is a frequency-domain estimator maximizing the likelihood of detecting a signal in noise with respect to the four amplitudes $\mathcal{A}^\mu$ \cite{Jaranowski1998}\footnote{Equations (\ref{eqn:h1})--(\ref{eqn:h4}) assume emission at $2 f_\star$ only, i.e. $\Phi(t) \approx 4 \pi f_\star t$, as appropriate for a perpendicular rotor. If the star's wobble angle is less than $90^{\circ}$, emission also occurs at $f_\star$, and eight amplitudes are involved, i.e. $h_1, \cdots, h_8$.}. It is defined as
\begin{equation}
\mathcal{F} = \frac{1}{2} x_\mu \mathcal{M}^{\mu \nu} x_\nu,
\end{equation}
where we write $x_\mu = (x|h_\mu)$, and $\mathcal{M}^{\mu \nu}$ denotes the inverse matrix of $\mathcal{M}_{\mu \nu}=(h_\mu|h_\nu)$. The scalar product $(\cdot|\cdot)$ featuring in the definitions of $x_\mu$ and $\mathcal{M}_{\mu \nu}$ is defined as a sum over single-detector scalar products,
\begin{eqnarray}
\nonumber (x|y) &=& \mathop{\sum} \limits_{X} (x^X|y^X) \\
        &=& \mathop{\sum} \limits_{X} 4\Re \int_{0}^{\infty}df \frac{\tilde{x}^X(f)\tilde{y}^{X*}(f)}{S^X(f)},
\end{eqnarray}
where $X$ indexes the detector, $S^X(f)$ is the one-sided noise spectral density of detector $X$, the tilde denotes a Fourier transform, and $\Re$ returns the real part of a complex number \cite{Prix2007}. 

The expectation value of the $\mathcal{F}$-statistic is 
\begin{equation}
E[2 \mathcal{F}]=4+\rho_0^2,
\end{equation}
where 
\begin{equation}
\rho_0 = (\mathcal{A}^\mu \mathcal{M}_{\mu \nu} \mathcal{A}^\nu)^{1/2}
\end{equation}
stands for the optimal SNR given a signal in Gaussian noise. The random variable $2\mathcal{F}$ is distributed according to a non-central chi-squared probability density function with four degrees of freedom, $p(2\mathcal{F})=\chi^2(2\mathcal{F}; 4,\rho_0^2)$, whose non-centrality parameter $\rho_0^2$ is related to the amplitudes by
\begin{equation}
\rho_0^2 = \frac{1}{2}[A(\mathcal{A}_1^2 + \mathcal{A}_3^2) + B(\mathcal{A}_2^2 + \mathcal{A}_4^2) + 2C(\mathcal{A}_1 \mathcal{A}_2 + \mathcal{A}_3 \mathcal{A}_4)],
\end{equation}
with $A=(a|a)$, $B=(b|b)$, and $C=(a|b)$. When there is no signal, the probability density function centralizes to give $p(2\mathcal{F})=\chi^2(2\mathcal{F}; 4,0)$. It can be shown that $\rho_0$ equals the SNR, with
\begin{equation}
\rho_0^2=\frac{K h_0^2T_\text{drift}}{S_n(2f_\star)},
\end{equation}
where the constant $K$ depends on the right ascension, declination, polarization and inclination angles of the source. Averaging without bias over these angles yields $K=4/25$ for a perpendicular rotor and an interferometer with perpendicular arms \cite{Jaranowski1998}.

\subsection{Detectability versus $h_0$}
\label{sec:isolated_tests}
We begin by illustrating the performance of the Viterbi algorithm with some representative examples. Seven sets of synthetic data are created for $T_\text{obs}=370$\,d at two detectors (H1 and L1) with $1\leq h_0/(10^{-26})\leq 20$ superposed on noise at a level typical of Advanced LIGO's design sensitivity, viz. ${S_n(2f_\star)}^{1/2}= 4 \times 10^{-24}$\,Hz$^{-1/2}$, near the instrument's most sensitive frequency \cite{Aasi2015,Messenger2015}. Sky position and source orientation are specified in Table \ref{tab:sig-params}. The synthetic SFTs are generated using \textit{Makefakedata} version 4 from the LIGO data analysis software suite LALApps\footnote{https://www.lsc-group.phys.uwm.edu/daswg/projects/lalsuite.html}. For each test, we take $T_\text{drift}=10$\,d, divide one year of data into $N_{T+1}=37$ segments, and create a 1-Hz band of $\mathcal{F}$-statistic output containing the injected $f_\star(t)$ for each segment. We take $\Pi_{q_i} = N_Q^{-1}=\Delta f_\text{drift}$ as the prior and $\ln L_{o(t),q_i}=\mathcal{F}$ from the $\mathcal{F}$-statistic output for each 10-day segment. We also use the transition probability matrix given by equation (\ref{eqn:trans_matrix}). For each data set, we calculate the root-mean-square deviation $\varepsilon$ (in Hz) between the optimal Viterbi path and the injected $f_\star(t)$. Tracking is deemed successful, if $\varepsilon$ is smaller than one $\mathcal{F}$-statistic frequency bin (width $\Delta f_\text{drift}$). A systematic Monte-Carlo calculation of the threshold for detection lies outside the scope of this paper.

\begin{table}
	\centering
	\setlength{\tabcolsep}{8pt}
	\begin{tabular}{lll}
		\hline
		\hline
		Parameter & Value & Units \\
		\hline
		$f_\star$ & 111.1 & Hz \\
		$\dot{f_\star}$ & 0.0 & Hz\,s$^{-1}$ \\
		$\psi$ & 4.08407 & rad\\
		$\cos\iota$ & 0.71934 & $-$ \\
		$\phi_0$ & 0.0 & rad\\
		$\alpha$ & 4.27570 & rad\\
		$\delta$ & $-$0.27297 & rad\\
		$S_n (2f_\star)^{1/2}$ & $4 \times 10^{-24}$ & Hz$^{-1/2}$ \\
		\hline
		\hline
	\end{tabular}
	\caption[parameters]{Injection parameters used to create the synthetic data analysed in Sections \ref{sec:isolated_tests} and \ref{sec:binary_tests}.}
	\label{tab:sig-params}
\end{table}

\begin{figure*}
	\centering
	\subfigure[]
	{
		\label{fig:iso-2e-25}
		\scalebox{0.26}{\includegraphics{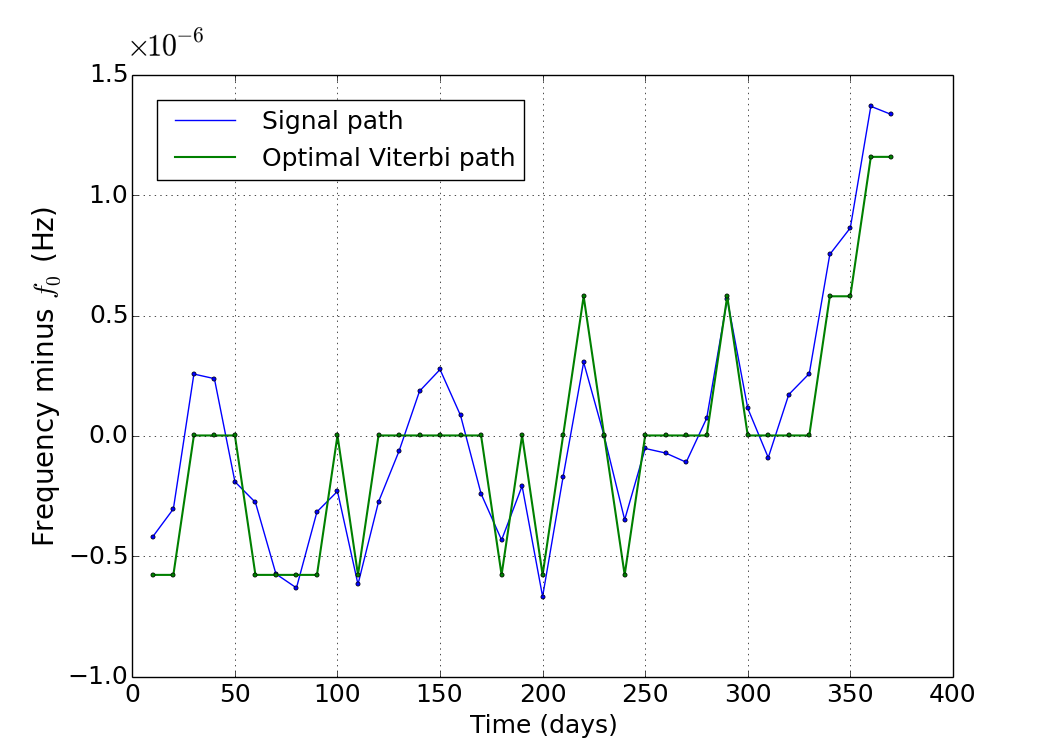}}
	}
	\subfigure[]
	{
		\label{fig:iso-1e-25}
		\scalebox{0.26}{\includegraphics{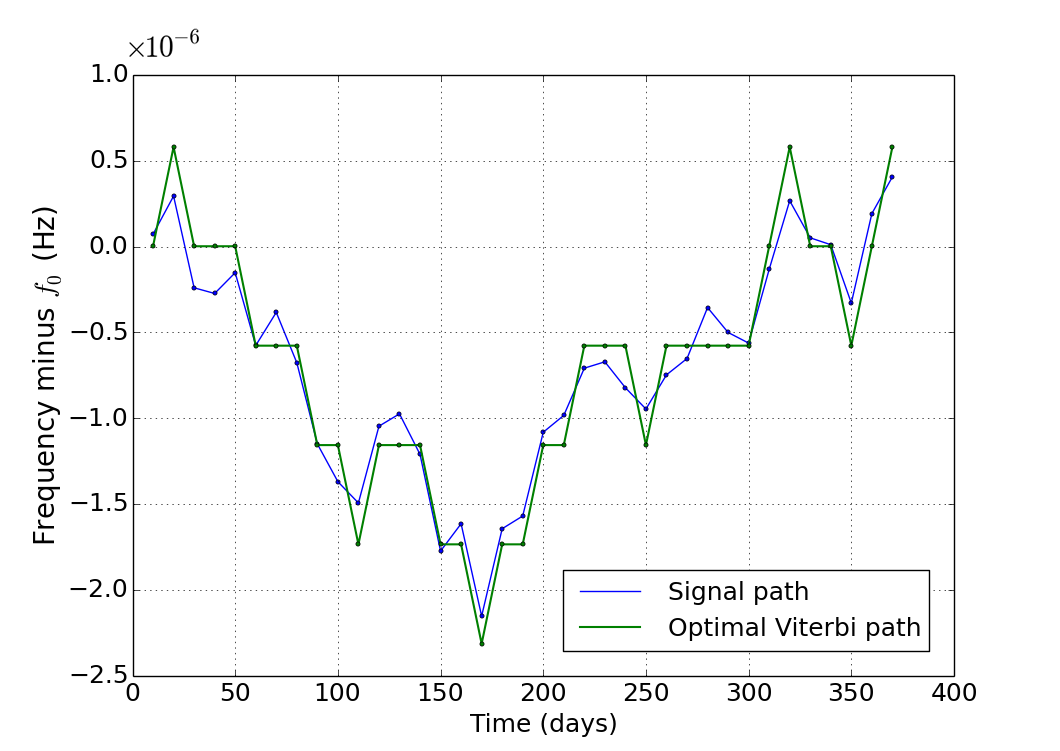}}
	}
	\subfigure[]
	{
		\label{fig:iso-8e-26}
		\scalebox{0.26}{\includegraphics{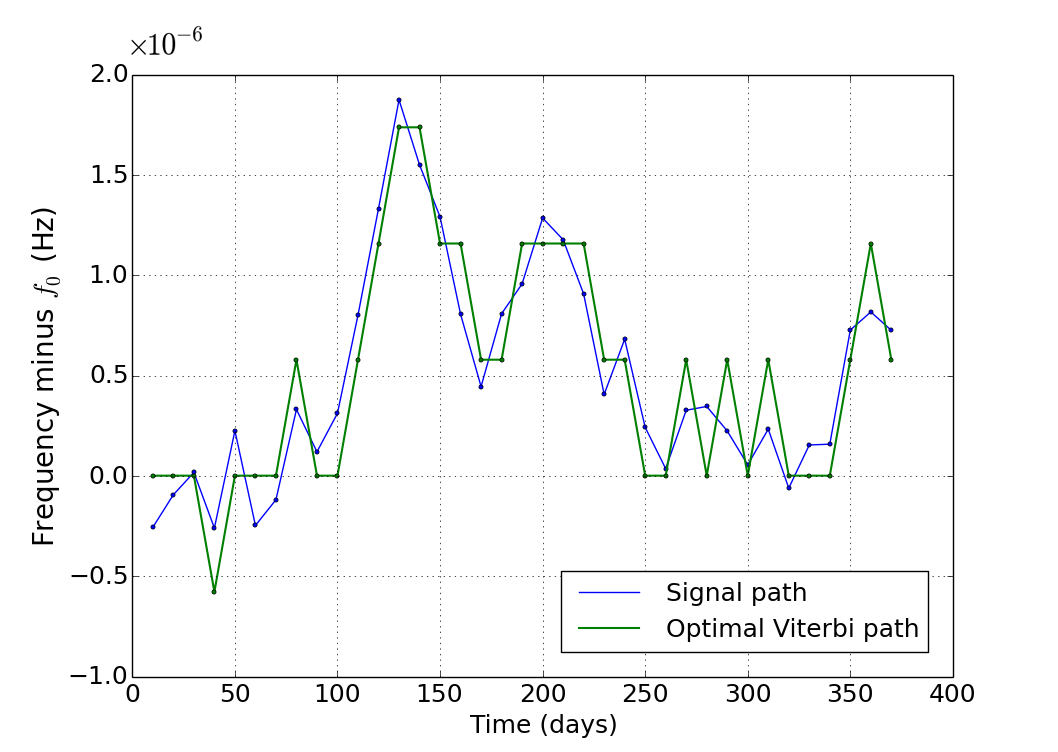}}
	}
	\subfigure[]
	{
		\label{fig:iso-6e-26}
		\scalebox{0.26}{\includegraphics{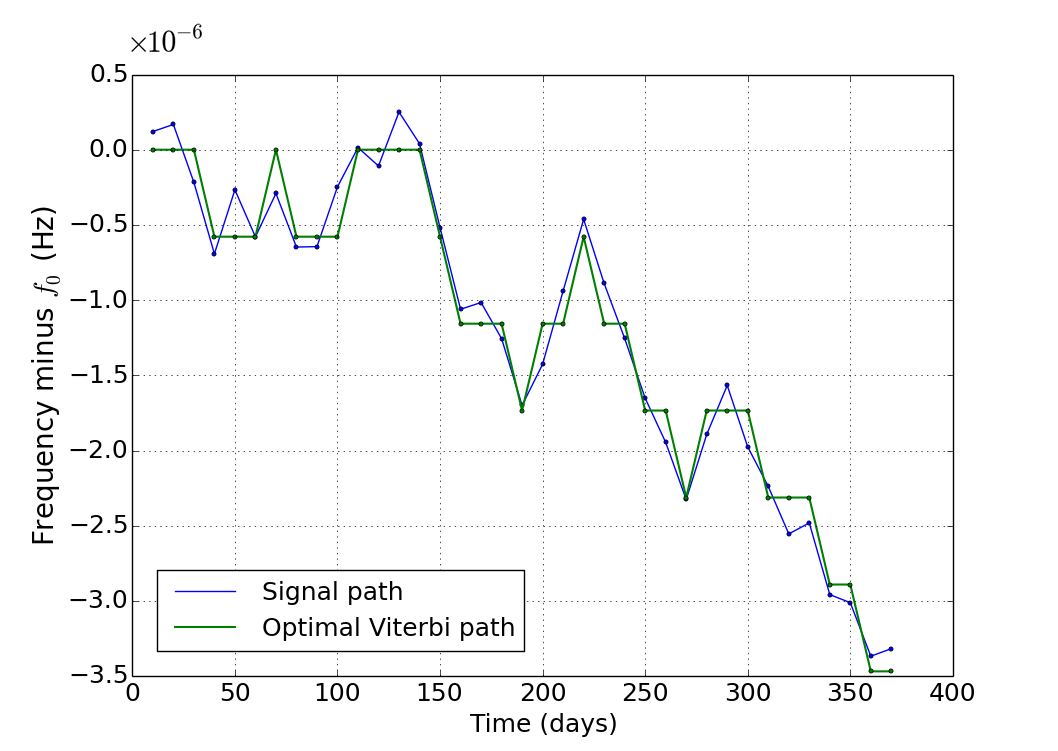}}
	}
	\subfigure[]
	{
		\label{fig:iso-4e-26}
		\scalebox{0.26}{\includegraphics{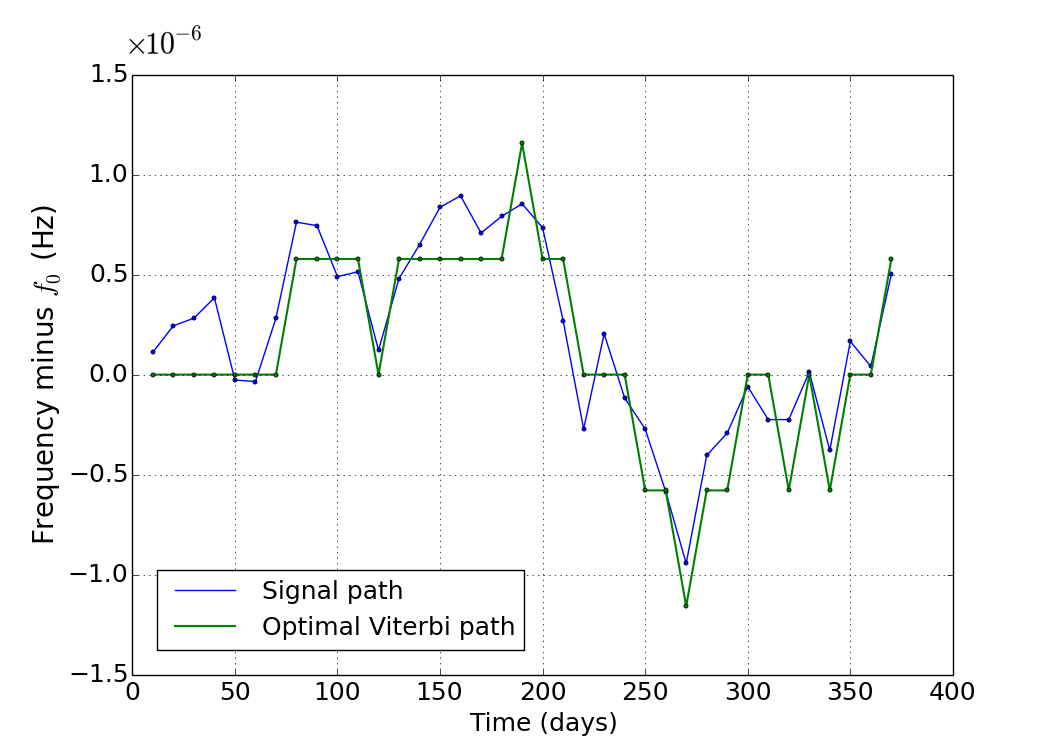}}
	}
	\subfigure[]
	{
		\label{fig:iso-2e-26}
		\scalebox{0.26}{\includegraphics{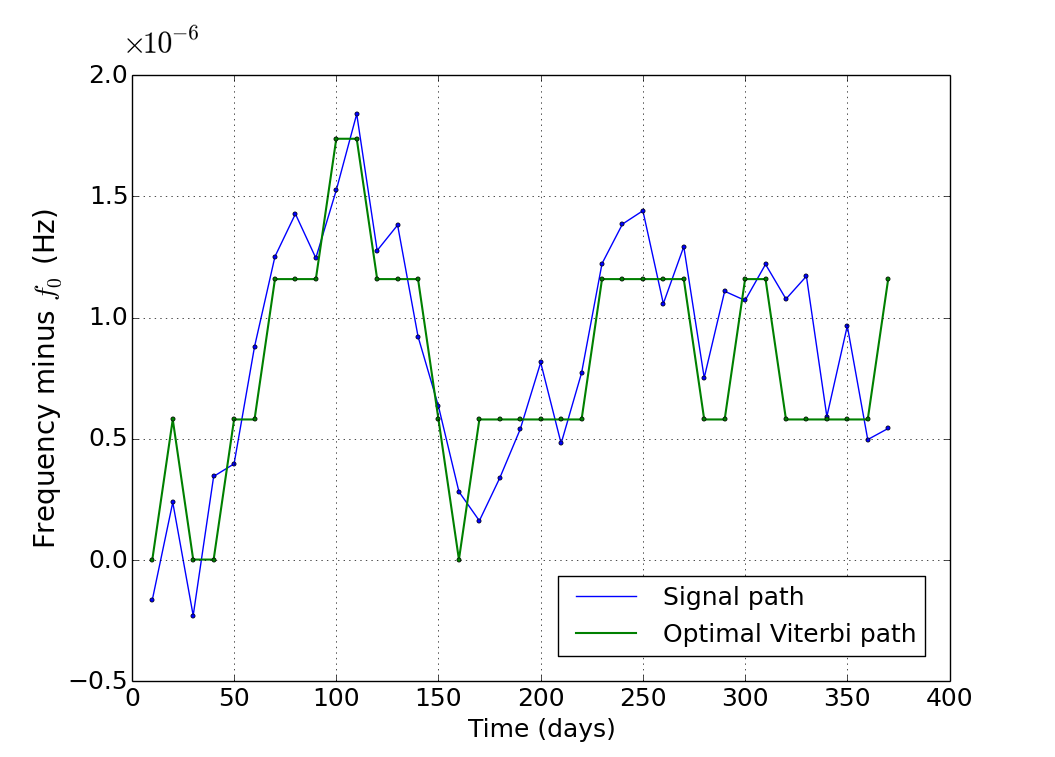}}
	}
	\subfigure[]
	{
		\label{fig:iso-1e-26}
		\scalebox{0.26}{\includegraphics{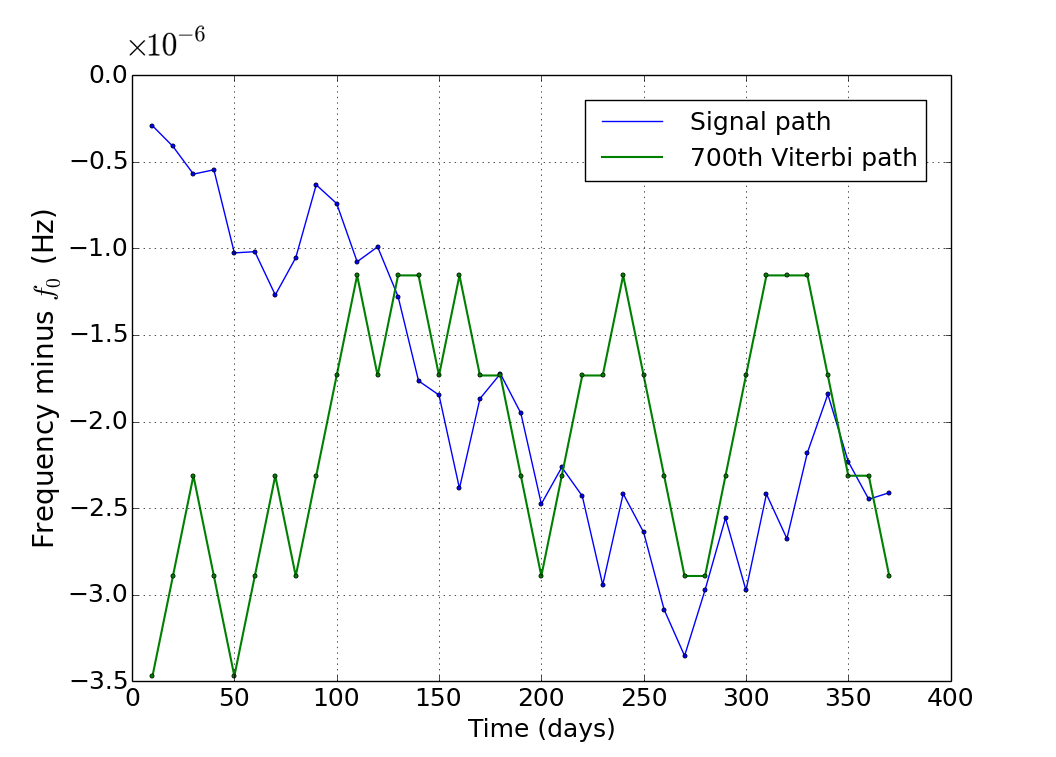}}
	}
	\caption{True $f_\star(t)$ (blue curve) and Viterbi path (green curve) for the seven injected signals in Table \ref{tab:results}, representing an isolated neutron star. Panels (a)--(f) display optimal Viterbi paths for $h_0/10^{-26}=20$, 10, 8, 6, 4, 2 respectively; a good match is obtained in each case. Panel (g) displays the closest-matching path (i.e. smallest $\varepsilon$) for $h_0=1\times 10^{-26}$; the match is poor. The units on the horizontal (time) and vertical (frequency) axes are days and Hz respectively. The symbol $f_0$ in the vertical axis label stands for $f_\star(0)$.}
	\label{fig:isolated_tracking_results}
\end{figure*}

The outcomes of the tests above are presented in Table \ref{tab:results} and Figure \ref{fig:isolated_tracking_results}. In Figures \ref{fig:iso-2e-25}--\ref{fig:iso-2e-26}, we see examples where the injected $f_\star(t)$ agrees closely with the optimal path reconstructed by the Viterbi algorithm. For $h_0/10^{-26}=20$, 10, 8, 6, 4, 2, the maximum root-mean-square error is $0.47 \Delta f_\text{drift}$. It arises mostly because the HMM takes one frequency bin as the smallest step, while the injected $f_\star(t)$ jumps to any value within $\pm 1$ bin. For $h_0 = 1 \times 10^{-26}$, the optimal Viterbi path is not a good match; the maximum error is $\varepsilon = 2 \times 10^5 \Delta f_\text{drift}$. Indeed the closest match to the injected signal is the 700-th Viterbi path [see Figure \ref{fig:iso-1e-26}], and even then the match is worse than that for all the tests with $h_0 \geq 2 \times10^{-26}$ ($\varepsilon = 2.15 \Delta f_\text{drift}$). 

\begin{figure*}
	\centering
	\subfigure[]
	{
		\label{fig:rmse-1}
		\scalebox{0.12}{\includegraphics{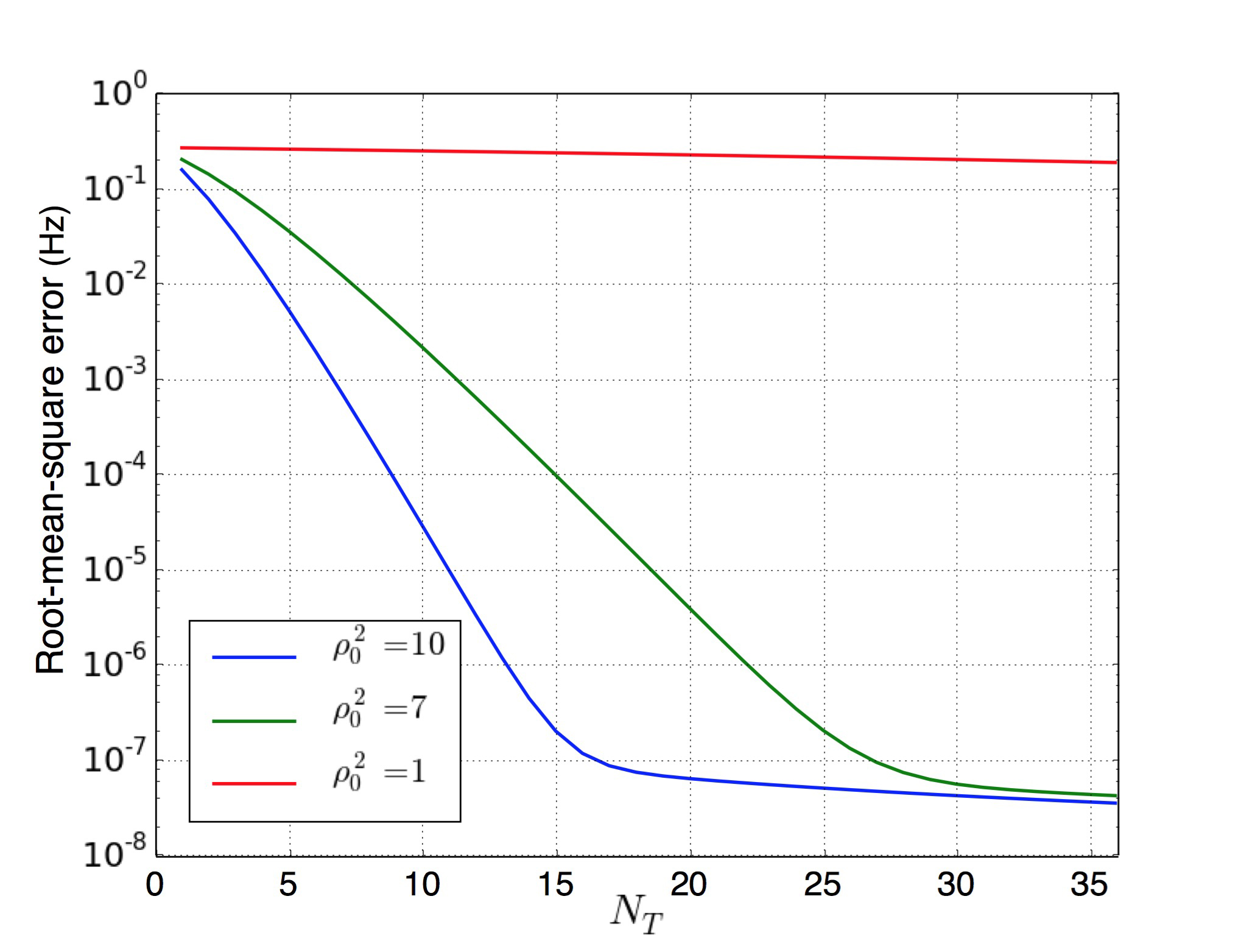}}
	}
	\subfigure[]
	{
		\label{fig:rmse-2}
		\scalebox{0.12}{\includegraphics{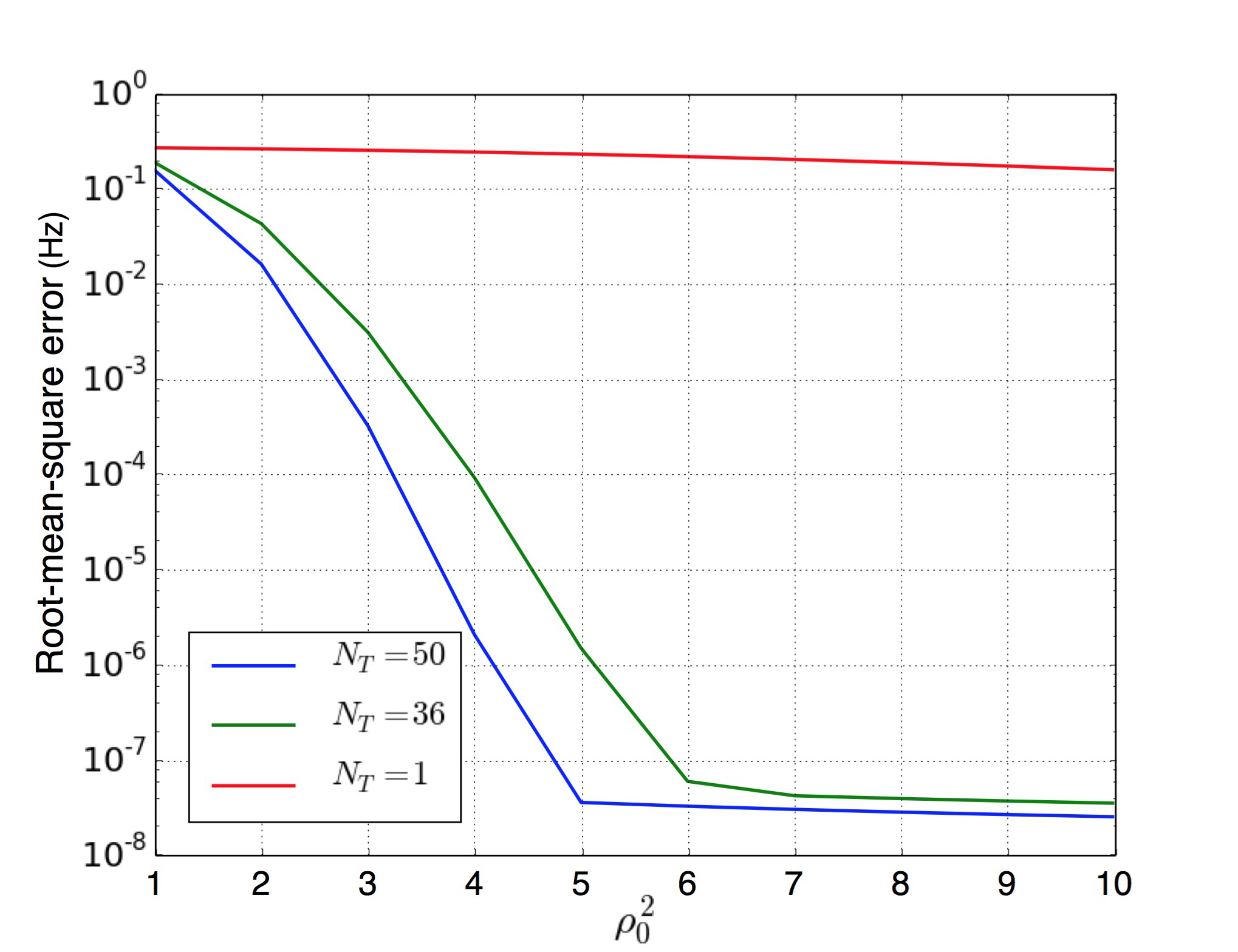}}
	}
	\caption{Formal root-mean-square error estimate in the vicinity of the detection threshold as a function of (a) $N_T$, with fixed $\rho_0^2=10$, 7 and 1, corresponding to $h_0=3.4 \times 10^{-26}$, $2.8 \times 10^{-26}$ and $1 \times 10^{-26}$, and (b) $\rho_0^2$, with fixed $N_T=50$, 36 and 1. The nearly horizontal, rightmost parts of the blue and green curves correspond to the CR lower bound \cite{Rife1974}.} 
	\label{fig:rmse}
\end{figure*}

The rapid loss of detectability experienced at $h_0 \approx 1 \times 10^{-26}$ in Table \ref{tab:results} is expected theoretically. Figure~\ref{fig:rmse} displays the formal root-mean-square error computed numerically as a function of $N_T$ for fixed $\rho_0^2$ [Figure \ref{fig:rmse-1}] and as a function of $\rho_0^2$ for fixed $N_T$ [Figure \ref{fig:rmse-2}]. The error is approximated as a linear combination of two terms, one arising from the probability of an outlier (see Sections \ref{sec:prob_distribution} and \ref{sec:prob_outlier}), and the other set by the Cram\'{e}r-Rao (CR) lower bound \cite{Rife1974}. For large SNR, the error variance of the estimator approaches the CR bound, i.e. the nearly horizontal, rightmost segments of the blue and green curves. As the SNR decreases, the error variance departs from the CR bound more and more and eventually becomes unbounded. For example, looking at the red curve in Figure \ref{fig:rmse-1} [for $\rho_0^2=1$, corresponding to the injection in Figure \ref{fig:iso-1e-26} with $h_0=1 \times 10^{-26}$], we see that the error variance does not asymptotically approach the CR bound over the plotted range of $N_T$. By contrast, in Figure \ref{fig:rmse-2} for example, the green curve (for $N_T=36$) hugs the CR bound for $\rho_0^2 \gtrsim 6$, corresponding to
$h_0 > 2.6 \times 10^{-26}$. However, as the SNR decreases in the regime $\rho_0^2 \lesssim 6$, the error diverges rapidly away from the turning point in the green curve,
with $\varepsilon \approx 10^{-0.5 \rho_0^3}$\,Hz as a rough approximation. This behaviour is typical at low SNR near the detection boundary \cite{Rife1974}. The probability that the optimal Viterbi path coincides with the injection in Figure \ref{fig:iso-1e-26} is approximately $35\%$, if the test is repeated for a large number of realizations of the noise. We quantify the probability that the optimal Viterbi path matches the injection in Section \ref{sec:prob_outlier}. Notice that, as always, extending $T_\text{obs}$ or increasing $T_\text{drift}$ (if the wandering is slow enough) improves the tracking.

\begin{table}
	\centering
	\setlength{\tabcolsep}{6pt}
	\begin{tabular}{llll}
		\hline
		\hline
		$h_0 (10^{-26})$ & Detect? & $\varepsilon$ (Hz)& $\varepsilon/\Delta f_\text{drift}$\\
		\hline
		$20.0$ & $\checkmark$ & $1.866 \times 10^{-7}$ & 0.322 \\
		$10.0$ & $\checkmark$ & $1.663\times 10^{-7}$ & 0.287 \\
		$8.0$ & $\checkmark$ & $2.132\times 10^{-7}$ & 0.368\\
		$6.0$ & $\checkmark$  & $1.683\times 10^{-7}$ & 0.291\\
		$4.0$ & $\checkmark$  & $2.054 \times 10^{-7}$ &0.355\\
		$2.0$ & $\checkmark$  & $2.719 \times 10^{-7}$& 0.470\\
		$1.0$ & $\times$  &0.112& $2 \times 10^5$ \\
		\hline
		\hline
	\end{tabular}
	\caption[detection results]{Outcome of Viterbi tracking for injected signals from isolated sources with the parameters in Table \ref{tab:sig-params}, $T_\text{obs}=370$\,d, $T_\text{drift}=10$\,d, and wave strain $h_0$. The root-mean-square error $\varepsilon$ between the optimal Viterbi track and injected $f_\star(t)$ is quoted in Hz and in units of $\Delta f_\text{drift}$, the $\mathcal{F}$-statistic frequency bin width.}
	\label{tab:results}
\end{table}

\subsection{Distribution of path probabilities}
\label{sec:prob_distribution}
The logarithm of equation (\ref{eqn:prob}) expresses $\ln P(Q|O)$ as a sum of log likelihoods, each of which is chi-squared-distributed with four degrees of freedom, if $G(f_\star)$ is the $\mathcal{F}$-statistic. As the chi-squared distribution is additive, we can easily calculate the probability density function of $z=P[Q^{(k)}|O^{(k)}]$ as a function of step number $k$, with $0 \leq k \leq N_T$, along any Viterbi path. If the Viterbi path coincides exactly with the true path, we obtain the probability density function
\begin{equation}
\label{eqn:chi-2-non-central}
p(z) = \chi^2(z;4k, k\rho_0^2). 
\end{equation}
If the Viterbi path does not intersect the true path anywhere, we obtain
\begin{equation}
\label{eqn:chi-2-central}
p(z) = \chi^2(z;4k, 0).
\end{equation}
If the Viterbi path intersects the true path at some steps but not others, $p(z)$ lies somewhere between $\chi^2(z;4k, k\rho_0^2)$ and $\chi^2(z;4k, 0)$. Note that $p(z)$ for the optimal Viterbi path lies somewhere between the above bounds but its functional form differs from (\ref{eqn:chi-2-non-central}) and (\ref{eqn:chi-2-central}). The Viterbi algorithm maximizes $z$ over all $3^k$ paths of length $k$ terminating at a given frequency bin $q(t_k)$. Hence, for pure noise, $p(z)$ for the optimal Viterbi path is constructed from (\ref{eqn:chi-2-central}) via the extreme value theorem modified to account for the fact that the paths overlap and are therefore correlated. This calculation is hard to do analytically and is postponed to future work.

\begin{figure}[h]
	\centering
	\scalebox{0.3}{\includegraphics{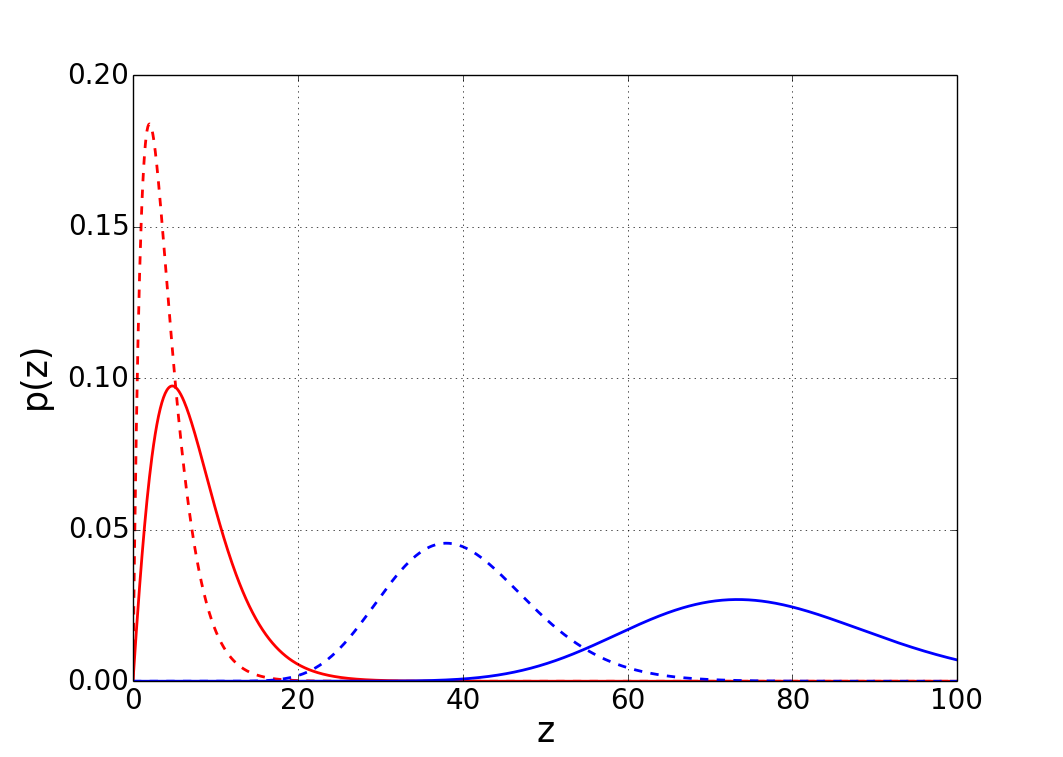}}
	\caption{Probability density function $p(z)$ for $z=P[Q^{(k)}|O^{(k)}]$ after $k=1$ (red curve) and $k=10$ (blue curve) Viterbi steps, assuming $Q^{(k)}$ intersects either perfectly (solid curve) or not at all (dashed curve) with the true path $f_\star(t)$. Parameters: $\rho_0^2=3.6$; corresponding to $h_0=2 \times 10^{-26}$ in Table \ref{tab:results}.} 
	\label{fig:chi2_pdf}
\end{figure}

Figure \ref{fig:chi2_pdf} displays $p(z)=\chi^2(z;4k, k\rho_0^2)$ (perfect intersection) and $ \chi^2(z;4k, 0)$ (no intersection) for the representative example $h_0=2 \times 10^{-26}$ (i.e. $\rho_0^2=3.6$) in Table \ref{tab:results}. The graph demonstrates clearly how it is progressively easier to detect a signal, as more Viterbi steps are taken. After $k=1$ steps, $\chi^2(z;4k, k\rho_0^2)$ and $ \chi^2(z;4k, 0)$ are hard to distinguish, with the difference confined to the tail. After $k=10$ steps, $\chi^2(z;4k, k\rho_0^2)$ and $ \chi^2(z;4k, 0)$ are well separated everywhere, including at the peak.

\subsection{Probability of no outlier}
\label{sec:prob_outlier}
In order for the optimal Viterbi path to be a reliable detection agent, we desire a high probability $1-P_\text{outlier}$ that the frequency bin containing the signal returns a higher value of $G(f)$ than all the other $N_Q-1$ frequency bins, which do not contain a signal, i.e. there are no outliers. Mathematically this translates to the condition $\delta_{q_i}(t_{N_T}) < \delta_{q_s}(t_{N_T})$ for all $i\neq s$ and for all possible values of the measurement $\delta_{q_s}(t_{N_T})$ in the state $q_s$ that contains the signal. From Section \ref{sec:prob_distribution} and the Optimality Principle we obtain
\begin{equation}
\label{eqn:no_outlier}
1-P_\text{outlier}=\int_0^\infty dz'\chi^2(z';4N_T,N_T \rho_0^2)P(z';4N_T,0)^{N_Q-1}, 
\end{equation}
where $P(z;\alpha, \beta) = \int_0^z dz' \chi^2(z';\alpha,\beta)$ is the chi-squared cumulative distribution function.

\begin{figure*}[htb]
	\centering
	\subfigure[]
	{
		\label{fig:no_outlier_1}
		\scalebox{0.32}{\includegraphics{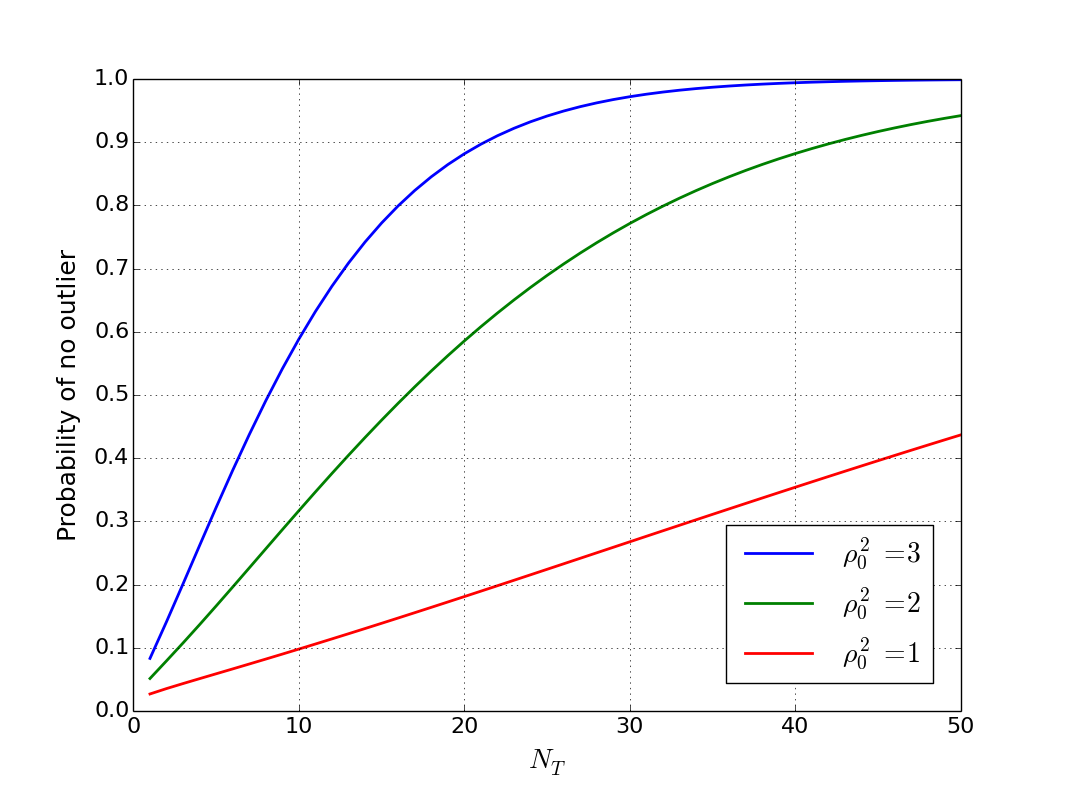}}
	}
	\subfigure[]
	{
		\label{fig:no_outlier_2}
		\scalebox{0.32}{\includegraphics{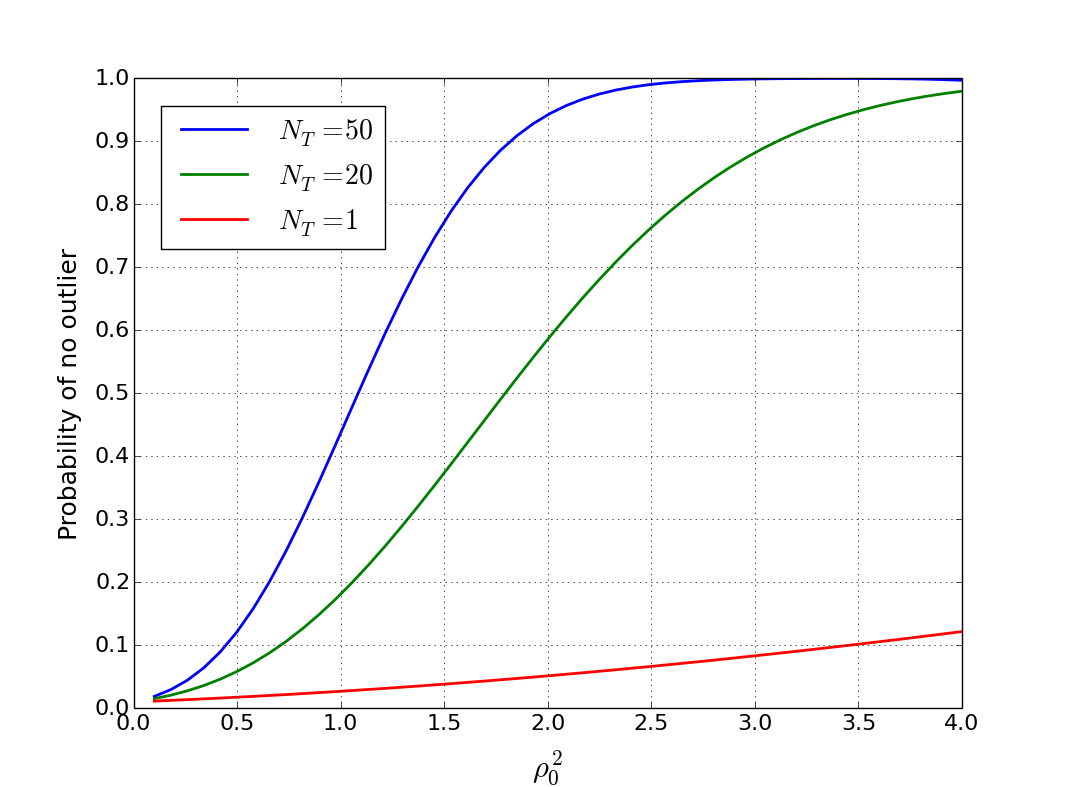}}
	}
	\caption{Probability of no outlier computed numerically with $N_Q=100$ $\mathcal{F}$-statistic frequency bins as a function of (a) $N_T$, with $\rho_0^2$ fixed, and (b) $\rho_0^2$, with $N_T$ fixed. Note that we have $N_Q \sim 10^6$ in real applications, but the integral in equation (\ref{eqn:no_outlier}) is hard to compute numerically in this regime. Fortunately the curve depends weakly on $N_Q$ for $N_Q \gg 1$, with $1-P_\text{outlier}$ varying by $\sim 4\%$ from $N_Q=100$ to $N_Q=200$.} 
	\label{fig:prob_no_outlier}
\end{figure*}

Figure \ref{fig:prob_no_outlier} shows the probability of no outlier computed numerically as a function of $N_T$ for fixed $\rho_0^2$ [Figure \ref{fig:no_outlier_1}] and as a function of $\rho_0^2$ for fixed $N_T$ [Figure \ref{fig:no_outlier_2}]. Detectability improves with the number of Viterbi steps. From the shape of the curves, it is clear that the lower the value of $\rho_0^2$ the more steps $N_T$ are required. 

\section{Binary Neutron Star}
\label{sec:binary}

\subsection{Matched filter: Bessel-weighted $\mathcal{F}$-statistic}
\label{sec:bessel}
When a biaxial rotor orbits a binary companion, the gravitational wave strain is frequency modulated due to the orbital Doppler effect. The signal at the detector is given by
\begin{equation}
x(t) = F_+(t)h_+(t) + F_\times(t) h_\times(t) + n(t),
\end{equation}
where $F_+$ and $F_\times$ are the beam-pattern functions defined in Equations (10) and (11) in Ref. \cite{Jaranowski1998}. For a Keplerian orbit, one has
\begin{equation}
\label{eqn:wave_strain}
h_{+,\times} (t)\propto \cos\{2 \pi f_\star[t+a_0\sin(2\pi t/P)]\},
\end{equation}
where $a_0$ is the projected semimajor axis, and $P$ is the orbital period. Expanding equation (\ref{eqn:wave_strain}) by the Jacobi-Anger identity~\cite{Abramowitz1964}, we obtain
\begin{equation}
\label{eqn:wave_strain_expansion}
h_{+, \times}(t) \propto \mathop{\sum} \limits_{n=-\infty}^{\infty} J_n(2 \pi f_\star a_0) \cos [2\pi (f_\star + n/P)t],
\end{equation}
where $J_n(z)$ is a Bessel function of order $n$ of the first kind.

The coefficients $J_n(z)$ in equation (\ref{eqn:wave_strain_expansion}) decay rapidly for $|n|>z$ ($z=2 \pi f_\star a_0\gg 1$), and the gravitational wave power is distributed into approximately $M= 2\text{ceil} (2 \pi f_\star a_0)+1$ orbital sidebands separated by $1/P$, where ceil$(x)$ denotes the smallest integer greater than or equal to $x$. 
Equation (\ref{eqn:wave_strain_expansion}) suggests that, over time intervals that are short compared to $T_\text{drift}$, the optimal matched filter takes the form of a convolution
\begin{equation}
\label{eqn:matched_filter}
G(f)=\mathcal{F}(f) \otimes B(f),
\end{equation}
where $B(f)$ is the squared modulus of the Fourier transform of the sum in equation (\ref{eqn:wave_strain_expansion}) heterodyned at $f_\star$, viz.
\begin{equation}
\label{eqn:matched_filter_cont}
B(f)=\sum\limits_{n=-(M-1)/2}^{(M-1)/2} [J_n(2 \pi f a_0)]^2 \delta(f-n/P).
\end{equation}

Let us now estimate approximately how the SNR depends on $\rho_0$, $f_\star$, and $a_0$. For the purpose of the following calculation we write $\mathcal{F}(f) \approx \rho_0^2 B(f) + W(f)$, where $W(f)$ is the $\mathcal{F}$-statistic of the noise, modeled as a chi-squared-distributed random variable with four degrees of freedom. This implies in particular $\text{var}(W)=8$, where var denotes the variance. The SNR yielded by $G(f)$ evaluated at the source frequency $f=f_\star$ is given by 
\begin{equation}
\text{SNR}=\frac{\rho_0^2 (B \otimes B)(f_\star)}{[\text{var}(W \otimes B)]^{1/2}}.
\end{equation}
For large $M$, $W \otimes B$ is approximately Gaussian, with 
variance 
\begin{equation}
\text{var}(W \otimes B) \approx 8\sum\limits_{n=-\infty}^\infty J_n^4(2 \pi f_\star a_0),
\end{equation}
and we also have 
\begin{equation}
(B \otimes B)(f_\star) = \sum\limits_{n=-\infty}^\infty J_n^4(2 \pi f_\star a_0), 
\end{equation}
implying
\begin{equation}
\text{SNR}=\frac{\rho_0^2}{2\sqrt{2}} \left[\sum\limits_{n=-\infty}^\infty J_n^4(2 \pi f_\star a_0)\right]^{1/2}.
\end{equation}
As $\sum\limits_{n=-\infty}^\infty J_n^4(z)$ is bounded by $z^{-1}$ for large $z$ \cite{Abramowitz1964}, we infer the lower bound
\begin{equation}
\label{eqn:snr_bessel}
\text{SNR} \ge \frac{\rho_0^2}{(16\pi f_\star a_0)^{1/2}}.
\end{equation} 
Note, this inequality requires $f_\star >40$ Hz, which is adequate for our purpose. 
In previous frequency domain searches for binaries, e.g. Sco X-1 \cite{Sammut2014}, the matched filter (\ref{eqn:matched_filter}) and (\ref{eqn:matched_filter_cont}) is replaced by an unweighted comb of orbital sidebands of the form
\begin{equation}
\label{eqn:comb}
B_\text{comb}(f)=\frac{1}{M}\sum\limits_{n=-(M-1)/2}^{(M-1)/2} \delta(f-n/P),
\end{equation}
called the $\mathcal{C}$-statistic. By an argument similar to the one in the previous paragraph, we have
\begin{equation}
\rho_0^2 (B \otimes B_\text{comb})(f_\star) = \rho_0^2, 
\end{equation}
\begin{equation}
\text{var}(W \otimes B_\text{comb}) = 32 \pi f_\star a_0,
\end{equation}
and hence
\begin{equation}
\label{eqn:snr_comb}
\text{SNR} \ge \frac{\rho_0^2}{(32 \pi f_\star a_0)^{1/2}}. 
\end{equation}
Equations (\ref{eqn:snr_bessel}) and (\ref{eqn:snr_comb}) demonstrate that the Bessel-weighted matched filter can recover approximately $\sqrt{2}$ times the SNR of the $\mathcal{C}$-statistic. Unlike for an isolated source, where the SNR depends only on $\rho_0$, the SNR for a binary source is also inversely proportional to $\sqrt{f_\star}$ (i.e. SNR $\propto h_0/\sqrt{f_\star}$), adversely affecting performance at higher frequencies.

\begin{figure*}
	\centering
	\subfigure[]
	{
		\label{whole_band}
		\scalebox{0.3}{\includegraphics{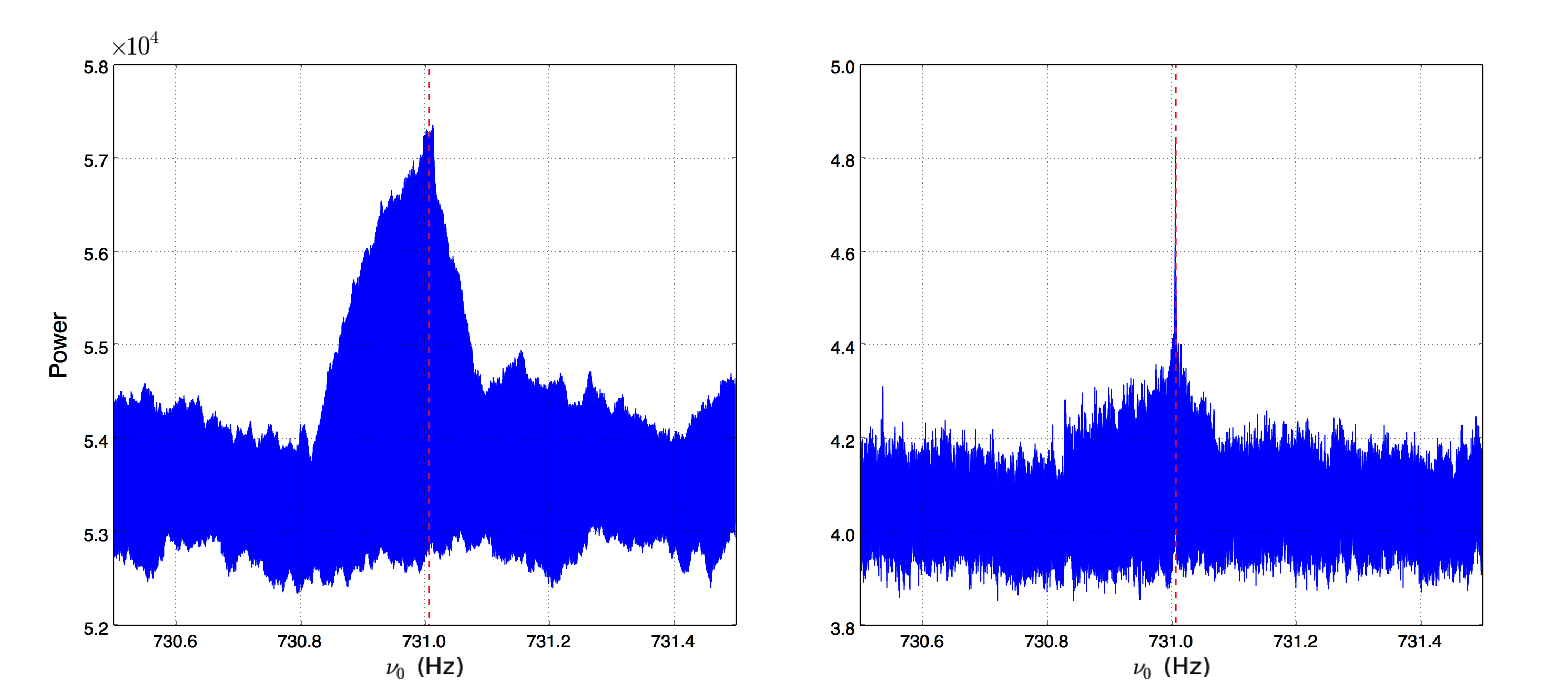}}
	}
	\subfigure[]
	{
		\label{zoom-in}
		\scalebox{0.26}{\includegraphics{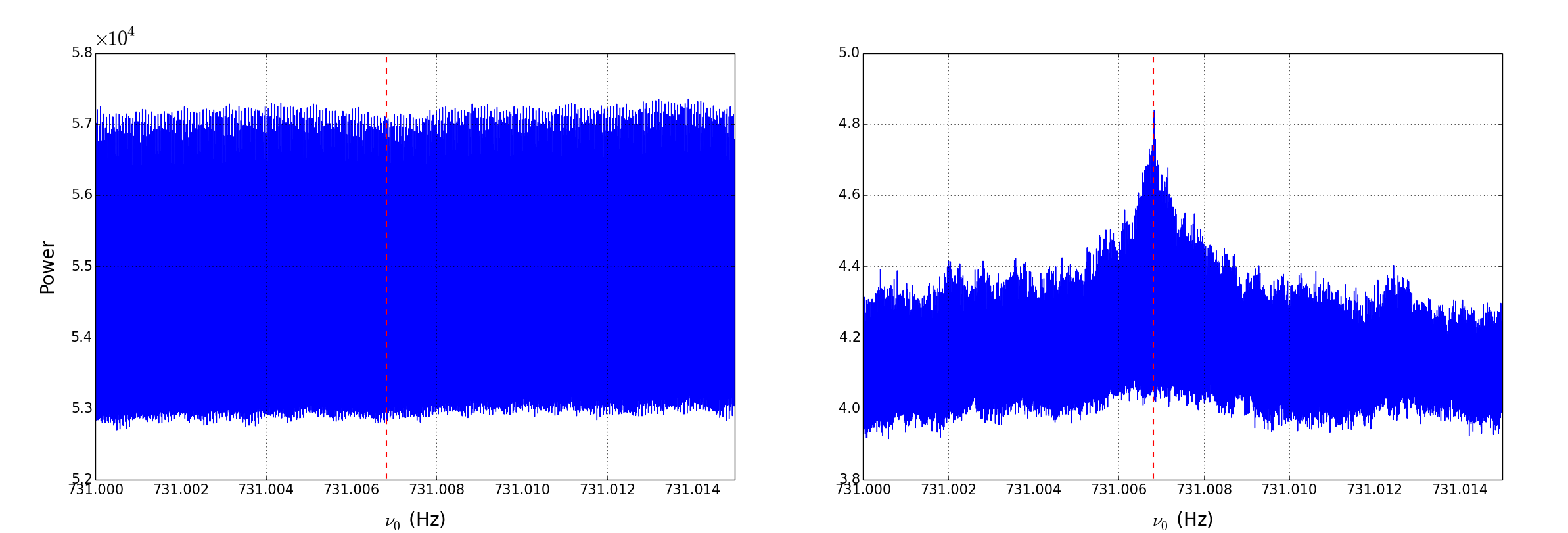}}
	}
	\caption{Convolution of the $\mathcal{F}$-statistic with a matched filter comprising unweighted (left panels) and Bessel-weighted (right panels) orbital sidebands for a 10-day data segment injected with a signal from a binary source. The red dashed lines indicate the injected frequency. The top and bottom panels show 1-Hz and 0.015-Hz bands around the injected frequency $f_\star(t)$ respectively.}
	\label{fig:bessel-vs-comb}
\end{figure*}

Figure \ref{fig:bessel-vs-comb} shows an example comparing the performance of the orbital sideband filters in equation (\ref{eqn:comb}) (left panels; unweighted) and (\ref{eqn:matched_filter_cont}) (right panels; Bessel weighted) on a 10-day data segment with an injected signal at $f_\star(t)=731.0068$\,Hz from a binary source. The Bessel-weighted filter takes $P=68023.7$\,s and $a\sin \iota=1.44$, i.e. values characteristic of Sco X-1 (see Sections \ref{sec:binary_tests} and \ref{sec:MDC}). Panels (a) and (b) (top and bottom) show 1-Hz and 0.015-Hz frequency bands containing the signal respectively. The injected frequency is marked by a red, vertical dashed line. Not only does the Bessel-weighted filter recover more signal power, but also the structure of its peak suits Viterbi tracking better. When we zoom into the peak [Figure \ref{zoom-in}; 731.0--731.015\,Hz], the $\mathcal{C}$-statistic output is relatively flat over a band of width $\approx 2M \Delta f_\text{drift}$, presenting the Viterbi tracker with multiple options, each with relatively low SNR. By contrast the Bessel-weighted filter marshals more of the power into a single, distinct peak, which is easier for the Viterbi algorithm to track. 

\subsection{Detectability versus $h_0$}
\label{sec:binary_tests}
\begin{table*}
	\centering
	\setlength{\tabcolsep}{6pt}
	\begin{tabular}{llll}
		\hline
		\hline
		Parameter & Value & Units& Description\\
		\hline
		$P$ & 68023.7 & s & Orbital period\\
		$a_0$ & 1.44 & s & Projected orbital semimajor axis\\
		$\Delta a_0$ & 0.18 & s & Measurement error in $a_0$\\
		$T_P$ & 1245984672&s & Time of periapsis passage in SSB\\
		$e$ & 0.0 & $-$&Orbital eccentricity\\
		\hline
		\hline
	\end{tabular}
	\caption{Orbital parameters used to create the synthetic data for the binary sources analysed in Section \ref{sec:binary_tests}. $2\Delta a_0$ is the width of the prior distribution of $a_0$.}
	\label{tab:bi-sig-params}
\end{table*}

\begin{table*}
	\centering
	\setlength{\tabcolsep}{8pt}
	\begin{tabular}{lllll}
		\hline
		\hline
		$h_0 (10^{-26})$ & Detect? & $\varepsilon$ (Hz)& $\varepsilon/\Delta f_\text{drift}$ & $\varepsilon P/M$\\
		\hline
		$20.0$ & $\checkmark$ & $3.296\times 10^{-7}$& 0.570 &$1.114\times 10^{-5}$\\
		$10.0$ & $\checkmark$ & $4.655\times 10^{-7}$& 0.804& $1.573\times 10^{-5}$ \\
		$8.0$ & $\checkmark$ & $4.709 \times 10^{-7}$&0.814&$1.591\times 10^{-5}$\\
		$6.0$ & $\times$  & 0.378 & $7\times 10^5$& 12.773\\
		\hline
		\hline
	\end{tabular}
	\caption{Outcome of Viterbi tracking for injected signals from binary sources with the parameters in Tables \ref{tab:sig-params} and \ref{tab:bi-sig-params}, $T_\text{obs}=370$\,d, $T_\text{drift}=10$\,d, and wave strain $h_0$. The root-mean-square error $\varepsilon$ between the optimal Viterbi track and injected $f_\star(t)$ is quoted in Hz, in units of $\Delta f_\text{drift}$, the $\mathcal{F}$-statistic frequency bin width, and in units of $M/P$, the half-width of the orbital sideband pattern.}
	\label{tab:bin_results}
\end{table*}

We begin by illustrating the performance of the Viterbi tracker for binary sources with some representative examples. We inject signals into Gaussian noise and generate synthetic SFTs for $T_\text{obs}=370$\,d at two interferometers using \textit{Makefakedata} version 4 as described in Section \ref{sec:isolated_ns}. We keep the same source parameters as in Section \ref{sec:isolated_tests} and introduce the orbital parameters listed in Table \ref{tab:bi-sig-params}, copied from Sco X-1 for definiteness. The analysis for each realisation proceeds in three steps. (1) We calculate the $\mathcal{F}$-statistic in a 1-Hz band containing the injection for each segment ($T_\text{drift}=10$\,d) of data and output 37 segments for a year. (2) We create a Bessel-weighted filter [equation (\ref{eqn:matched_filter_cont})] with $P=68023.7$\,s and $a\sin \iota=1.44$\,s, process each 10-day $\mathcal{F}$-statistic segment, and generate 37 $G(f)$ outputs. (3) We apply the Viterbi tracker to the $G(f)$ output and find the optimal Viterbi path like in Section \ref{sec:isolated_tests}.

In realistic applications, $a_0$ is often known approximately but not exactly from electromagnetic observations (typical uncertainty $\approx \pm 10 \%$) \cite{Galloway2014,Messenger2015}. Hence in general we track $a_0$ as well as $f_\star(t)$, as described in Sections \ref{sec:jump_prob} and \ref{sec:setup_markov}. However, for the tests in this section, our initial guess for $a_0$ matches exactly the injected value in Table \ref{tab:bi-sig-params}, so we only track $f_\star(t)$. (In Section \ref{sec:MDC}, $a_0$ is tracked too.)

The Bessel-weighted filter in equation (\ref{eqn:matched_filter_cont}) depends on $2 \pi f a_0$, complicating step (2) in the paragraphs above. Strictly speaking, the filter takes a slightly different form in each $\mathcal{F}$-statistic frequency bin within each 1-Hz band, which would be prohibitive computationally to implement. Instead, we execute step (2) using a filter with Bessel weightings $J_n(2 \pi \overline{f} a_0)$, where $\overline{f}$ is the central frequency in each 1-Hz band. The fractional error thereby introduced across the 1-Hz band is minimal ($\lesssim 1 \%$) compared to the fractional uncertainty in $a_0$ from electromagnetic observations. A similar approach was adopted in previous $\mathcal{C}$-statistic searches \cite{Sammut2014,SCO-X1-2015,Messenger2015}. The fractional uncertainty in $P$ from electromagnetic observations is typically a few parts in $10^7$ and can be neglected.

\begin{figure*}
	\centering
	\subfigure[]
	{
		\label{fig:bin-2e-25}
		\scalebox{0.3}{\includegraphics{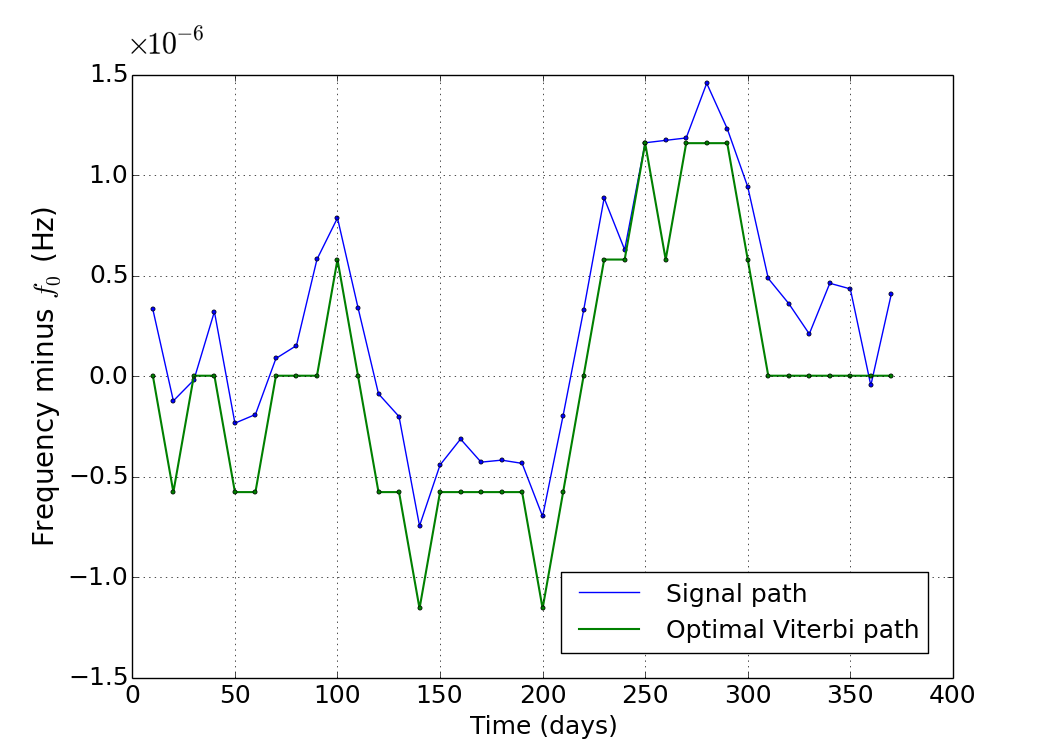}}
	}
	\subfigure[]
	{
		\label{fig:bin-1e-25}
		\scalebox{0.3}{\includegraphics{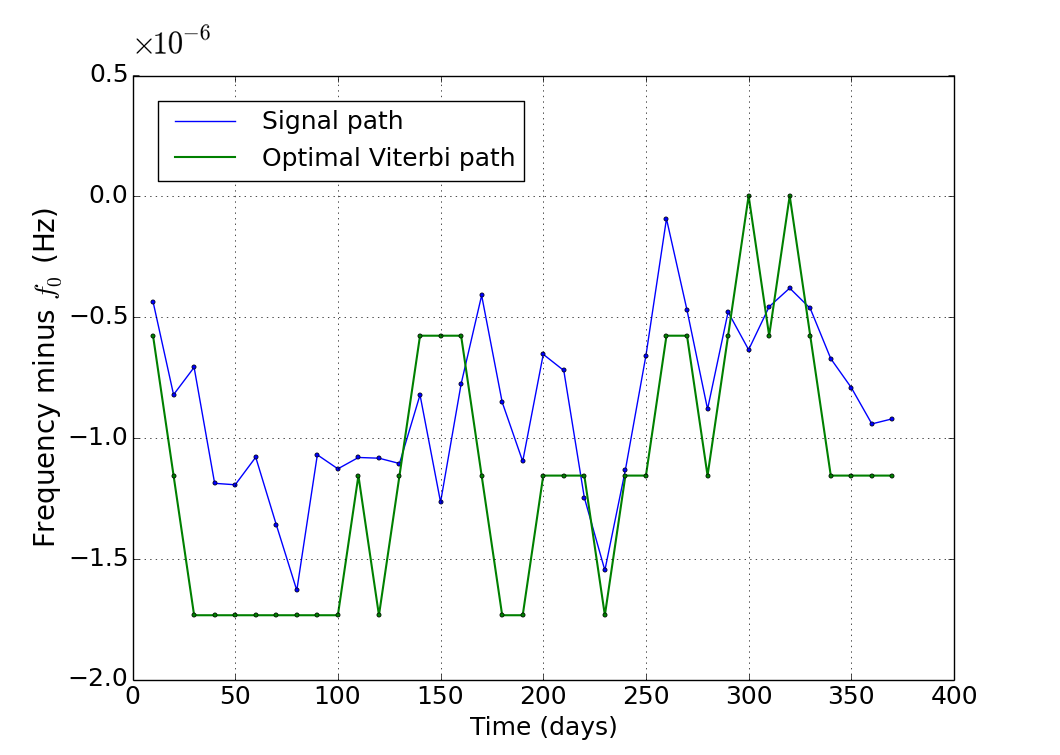}}
	}
	\subfigure[]
	{
		\label{fig:bin-8e-26}
		\scalebox{0.3}{\includegraphics{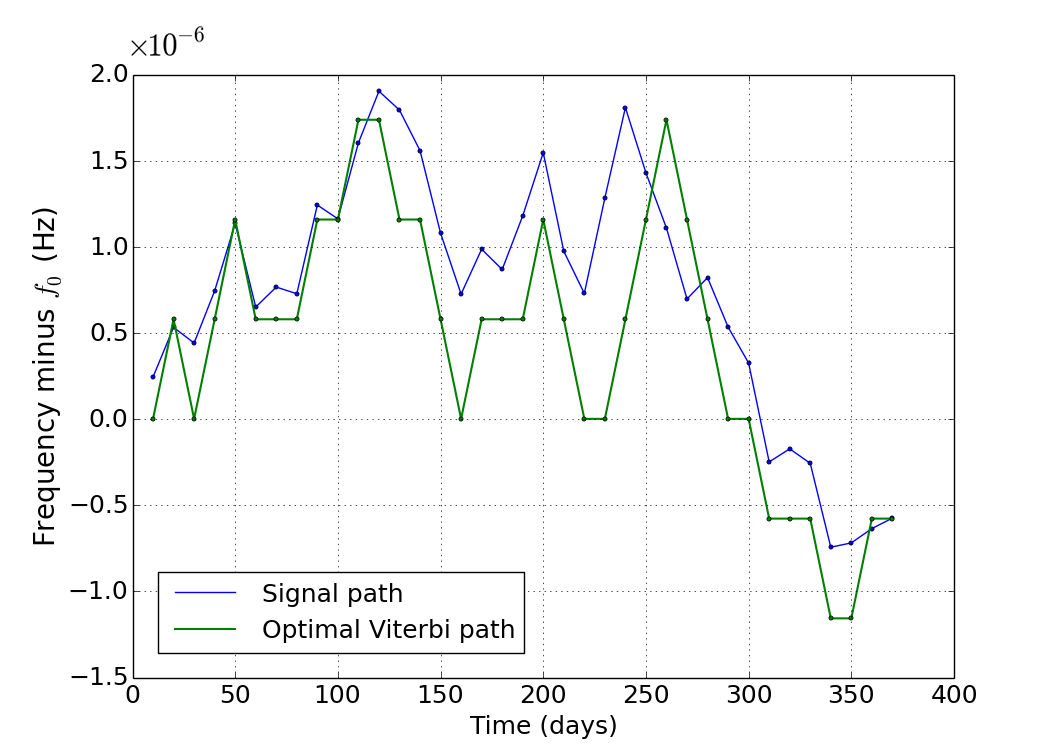}}
	}
	\subfigure[]
	{
		\label{fig:bin-6e-26}
		\scalebox{0.3}{\includegraphics{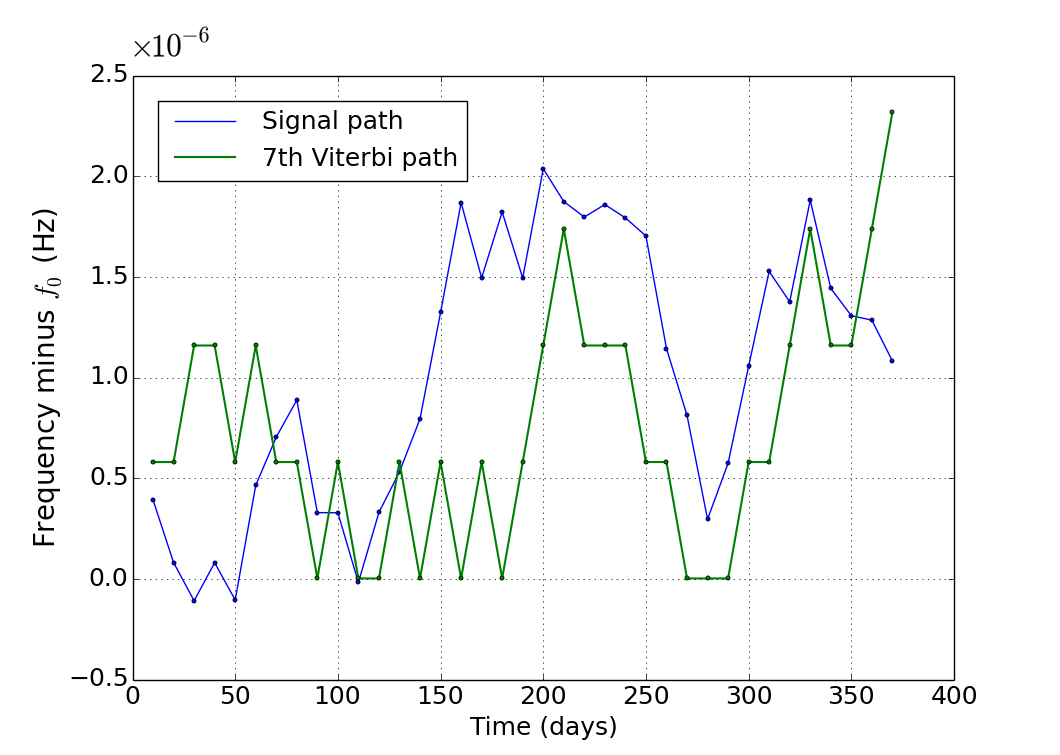}}
	}
	\caption{True $f_\star(t)$ (blue curve) and Viterbi path (green curve) for the four injected signals in Table \ref{tab:bin_results}, representing a neutron star in a binary orbit. Panels (a)--(c) display optimal Viterbi paths for $h_0/10^{-26}=20$, 10, 8 respectively; a good match is obtained in each case. Panel (d) is the closest-matching path (i.e. smallest $\varepsilon$) for $h_0=6\times 10^{-26}$. The units on the horizontal (time) and vertical (frequency) axes are days and Hz respectively. The symbol $f_0$ in the vertical axis label stands for $f_\star(0)$.}
	\label{fig:binary_drift}
\end{figure*}

The test outcomes are presented in Table \ref{tab:bin_results} and Figure \ref{fig:binary_drift}. Figures \ref{fig:bin-2e-25}--\ref{fig:bin-6e-26} show the tracking results for $h_0/10^{-26} = 20$, 10, 8, 6. Tracking is deemed successful, if the root-mean-square discrepancy $\varepsilon$ between the optimal Viterbi path and injected $f_\star(t)$ is less than one $G(f)$ frequency bin (width $\Delta f_\text{drift}$) or the width of the orbital sideband pattern ($M/P$), whichever is larger. In Figure \ref{fig:bin-2e-25}--\ref{fig:bin-8e-26}, the injected $f_\star(t)$ agrees well with the optimal Viterbi path. For $h_0/10^{-26} = 20$, 10, 8, the maximum root-mean-square error is 0.814 $\Delta f_\text{drift}$. For $h_0=6 \times 10^{-26}$, the optimal Viterbi path is a poor match with $\varepsilon= 7 \times 10^5 \Delta f_\text{drift}$. The closest match to the injected signal is the seventh Viterbi path with $\varepsilon=7.677 \times 10^{-7}\,\text{Hz} = 1.327 \Delta f_\text{drift}$ [see Figure \ref{fig:bin-6e-26}]. In other words, the sensitivity drops four-fold from $h_0 \approx 2 \times 10^{-26}$ for an isolated source to $h_0 \approx 8 \times 10^{-26}$ for a binary source.

\begin{figure}
	\centering
	\scalebox{0.2}{\includegraphics{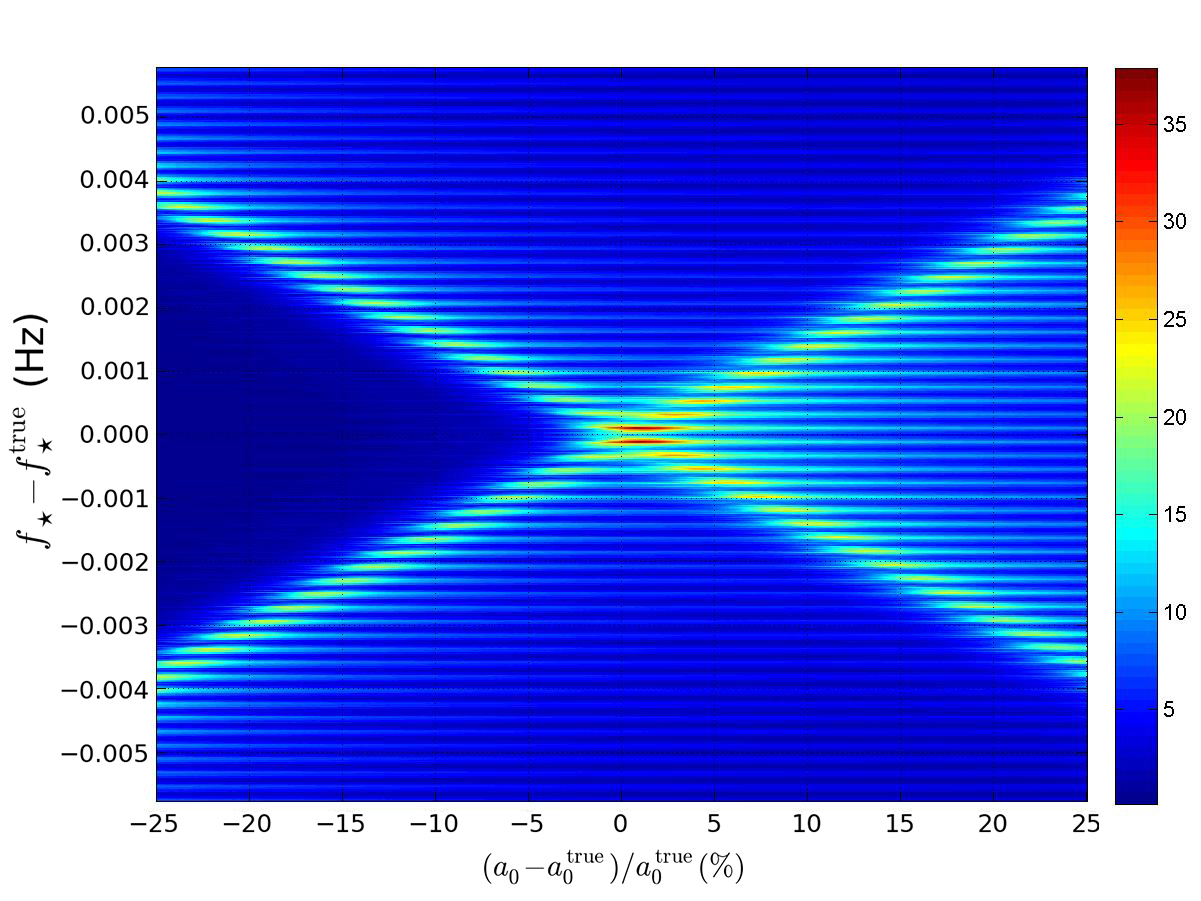}}
	\caption[log likelihood 2D]{Contour plot of the log likelihood $\ln P[Q^\ast(O)|O]$ as a function of $a_0$ (expressed as the percentage offset from the true, injected value $a_0^{\rm true}$) and $f_\star$ (expressed as the absolute offset from the true, injected value $f_\star^{\rm true}$) for a strong (SNR $\gg 1$) binary-star signal with constant $f_\star$ observed during a single 10-d segment ($T_{\rm obs}=10\,{\rm d}$). The bright colors (cyan, yellow, red) stand for the highest log likelihoods. The injected signal parameters are $f_\star^{\rm true}=111.1$\,Hz, $a_0^{\rm true}=1.44$, and $\Delta a_0=0.25 a_0^{\rm true}$, characteristic of Scorpius X-1.}
	\label{fig:loglik_2D}
\end{figure}

We quantify the error in tracking $f_\star$ as a function of the error in the assumed value of $a_0$, by injecting a strong signal with $h_0/\sqrt{S_n(2f_\star)}=10^7$\,Hz$^{1/2}$ into a 10-day segment, and tracking $f_\star$ as well as $a_0$ over 250 bins spanning $\pm 25 \%$ of the injected value $a_0^{\rm true}$. The results are displayed in Figure \ref{fig:loglik_2D}, a contour plot displaying the log likelihood $\ln P[Q^\ast(O)|O]$ as a function of $a_0$ (expressed as the percentage offset from the true, injected value $a_0^{\rm true}$) and $f_\star$ (expressed as the absolute offset from the true, injected value $f_\star^{\rm true}$). The bright colors (cyan, yellow, red) stand for the highest log likelihoods. The maximum value (in red) is found at the injected values of $f_\star^{\rm true}$ and $a_0^{\rm true}$. Caused by the $\pm 25 \%$ uncertainty in $a_0$, the maximum absolute offset of $f_\star$ recovered by the tracker from $f_\star^{\rm true}$ is $\varepsilon_{f_\star} \approx 0.004$\,Hz ($0.135\,M/P$). In reality the $\pm 10 \%$ uncertainty in $a_0$ leads to an uncertainty in $f_\star$, given by $\varepsilon_{f_\star} \approx 0.001$\,Hz ($0.003\,M/P$).

\section{Sco X-1 Mock Data Challenge: a ``realistic" example}
\label{sec:MDC}
In this section, we combine the Viterbi tracker and Bessel-weighted matched filter validated in Section \ref{sec:binary} to search for the 50 Sco-X-1-type signals (50--1500\,Hz) generated for the Sco X-1 Mock Data Challenge (Version 6) \cite{Messenger2015}. The aims of the exercise are three-fold: (i) to test the performance of the Viterbi tracker under ``realistic" conditions on a data set generated by an independent party; (ii) to compare the performance of the Viterbi tracker against the CrossCorr \cite{Dhurandhar2008,Chung2011,Whelan2015}, TwoSpect \cite{TwoSpectGoetz2011}, Radiometer \cite{RadiometerBallmer2006,Radiometer2007,Radiometer2011}, Sideband \cite{Sideband-Messenger2007,Sammut2014} and Polynomial \cite{Polynomial-Putten2010} pipelines which competed in the Mock Data Challenge; and (iii) to prepare the Viterbi tracker for Advanced LIGO observations. 

The parameters of the 50 injected signals in the Mock Data Challenge are listed in Table III in Ref. \cite{Messenger2015}. The 50 signals were originally ``closed", i.e. their parameters were kept secret, in order to compare blindly the competing pipelines from the perspectives of sensitivity, parameter estimation and efficiency. Four pipelines (TwoSpect, Radiometer, Sideband and Polynomial) competed under closed conditions in Ref. \cite{Messenger2015}. CrossCorr analysed the data in self-blinded mode, after the injection parameters were revealed. We note that $f_\star(t)$ does not wander for any of the injected signals, a situation which the Viterbi tracker with the tridiagonal transition matrix in Equation (\ref{eqn:trans_matrix}) handles easily and without bias. To mimic a real search, we claim a detection if the log likelihood of the optimal Viterbi path, $\ln P[Q^\ast(O)|O]$, exceeds its mean value plus seven standard deviations. The choice of seven standard deviations in this paper is arbitrary but it is broadly consistent with the thresholds chosen in the previous Sideband searches for Sco X-1 in LIGO S5 data \cite{Scox1-2015} and in Stage I of the MDC \cite{Messenger2015}, yielding approximately the same detection rate using a 10-day segment. A more systematic Monte-Carlo calculation of the threshold and false alarm rate lies outside the scope of this paper.

We conduct the search in three stages. Firstly, we pick the same 10-day segment of MDC data analysed in Ref. \cite{Messenger2015} by the Sideband pipeline, starting at GPS time 1245000000. We find that 12 out of 50 signals are detected, matching the performance of the $\mathcal{C}$-statistic. Secondly, for the 38 signals that are not detected in 10 days, we analyse a one-year stretch of data starting at GPS 1230338490. We find that 23 extra signals are detected, leaving 15 out of 50 undetected. The first two stages are performed with data from two interferometers due to computational limitations. In the third stage, we reanalyse the 15 remaining signals using three interferometers. Gratifyingly, we find that we detect six extra signals. Stages two and three are performed for $T_\text{obs}=1$\,yr, i.e. on the same footing as the four non-Sideband algorithms competing in Ref. \cite{Messenger2015}. We present the results from the three stages in detail in Section \ref{sec:10d-MDC}, \ref{sec:1yr-2ifo-MDC} and \ref{sec:1yr-3ifo-MDC} below and tabulate them in Table \ref{tab:MDC}.

We find that the error in the estimates of $f_\star(0)$ and $a_0$, denoted by $\varepsilon_{f_\star(0)}$ and $\varepsilon_{a_0}$ respectively, satisfy $\varepsilon_{f_\star(0)} < 7.4 \times 10^3 \Delta f_\text{drift}$ (i.e. $4.26\times10^{-3}$\,Hz) and $\varepsilon_{a_0} < 0.4 a_0$ in all cases where there is a successful detection.

\subsection{$T_\text{obs}=10$\,d,  two interferometers}
\label{sec:10d-MDC}
In the first stage, we pick the same 10-day segment of MDC data analysed in Ref. \cite{Messenger2015} by the sideband pipeline from two interferometers (H1 and L1), starting at GPS time 1245000000. We search a 1-Hz frequency band containing the signal for each injection, setting $P=68023.7$\,s and tracking a 0.72-s band of $a_0$ centred on the electromagnetic observation value 1.44\,s. A uniform prior is set for both $f_\star$ and $a_0$. The first stage successfully detects injections 1, 3, 15, 20, 32, 35, 59, 62, 65, 66, 75 and 84. We detect 12 signals rather than the 16 found by the Sideband pipeline in Ref. \cite{Messenger2015}, which used data from three interferometers. As a cross-check, we perform a supplementary search for the four missing signals with three interferometers and $T_{\rm obs} = 10\,{\rm d}$ and detect them all.

\subsection{$T_\text{obs}=1$\,yr, two interferometers}
\label{sec:1yr-2ifo-MDC}
For the 38 out of 50 signals that are not detected in a single 10-day segment, we do Viterbi tracking for $T_\text{obs} = 1$\,yr using data from two interferometers (H1 and L1). The search space and prior are the same as those in Section \ref{sec:10d-MDC}. In this stage 23 extra injections are successfully detected: 2, 5, 11, 14, 17, 19, 23, 26, 29, 36, 44, 47, 51, 60, 61, 67, 68, 76, 79, 83, 85, 95, and 98.

\subsection{$T_\text{obs}=1$\,yr, three interferometers}
\label{sec:1yr-3ifo-MDC}
For the remaining 15 signals that are not detected in the first two stages, we do Viterbi tracking for $T_\text{obs} = 1$\,yr using data from three interferometers (H1, L1 and V1). The search space and prior are the same as those in Sections \ref{sec:10d-MDC} and \ref{sec:1yr-2ifo-MDC}. In this last stage six out of the remaining 15 injections are successfully detected: 21, 50, 52, 54, 58, and 71.

\begin{figure}
	\centering
	\scalebox{0.25}{\includegraphics{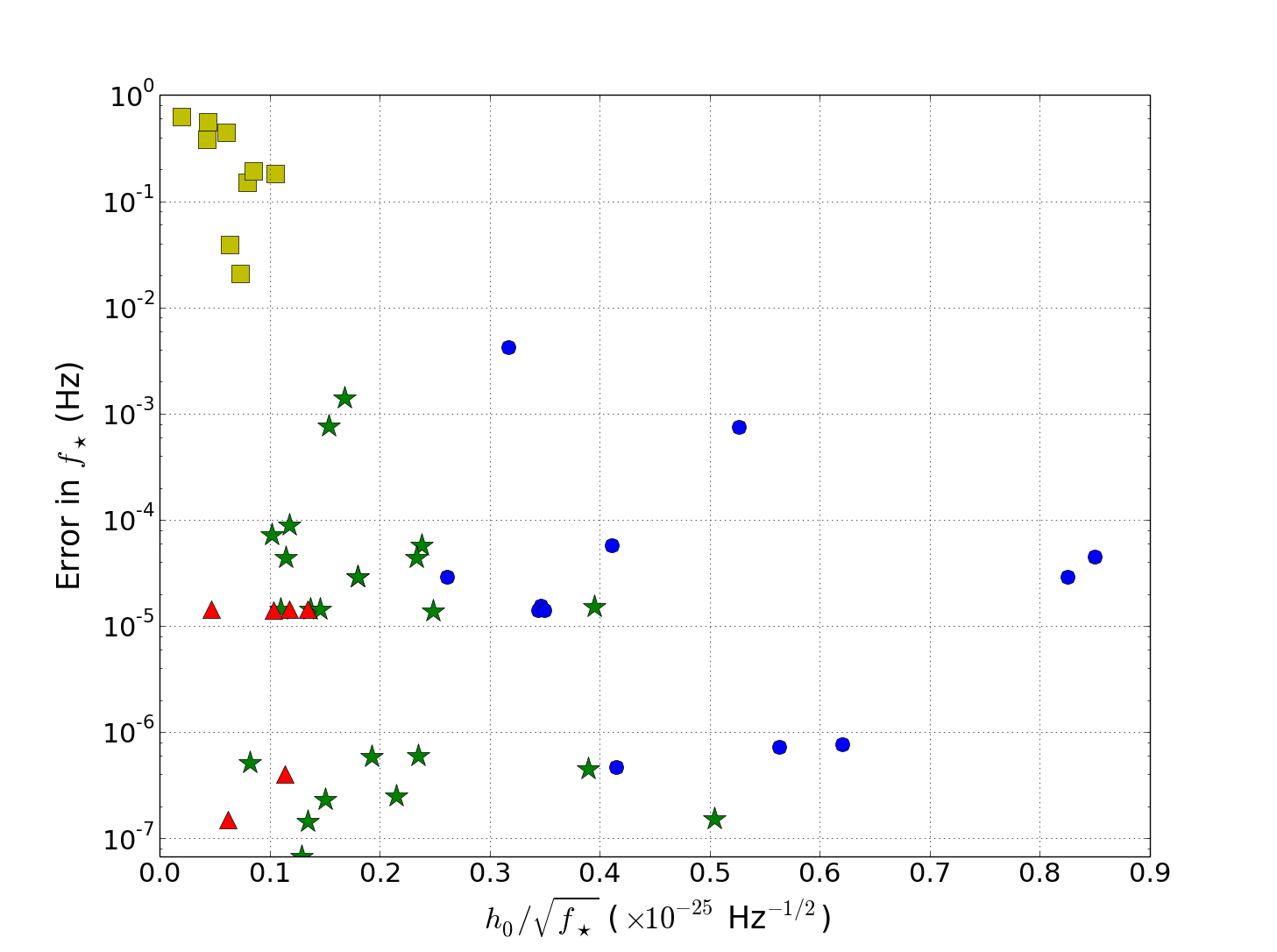}}
	\caption{Performance of the Viterbi tracker in Stage I of the Sco X-1 MDC. Error in estimated $f_\star$ as a function of $h_0/\sqrt{f_\star}$ for the 50 injected signals in Stage I of the Sco X-1 MDC. The blue circles, green stars and red triangles mark injections detected in stages one ($T_\text{obs}=10$\,d,  two interferometers), two ($T_\text{obs}=1$\,yr,  two interferometers) and three ($T_\text{obs}=1$\,yr,  three interferometers) respectively. The yellow squares mark the injections not detected in any of the three stages. Injection parameters are provided in Table \ref{tab:MDC}. }
	\label{fig:mdc_err}
\end{figure}

Figure \ref{fig:mdc_err} shows the error in estimated $f_\star$ as a function of $h_0/\sqrt{f_\star}$ for the 50 injected signals in Stage I of the Sco X-1 MDC. The circles, stars and triangles mark injections detected in stages one ($T_\text{obs}=10$\,d,  two interferometers), two ($T_\text{obs}=1$\,yr,  two interferometers) and three ($T_\text{obs}=1$\,yr,  three interferometers) respectively. The squares mark the injections not detected in any of the three stages. All the nine undetected signals have low SNR, with $h_0/\sqrt{f_\star} \lesssim 1 \times 10^{-26}$\,Hz$^{-1/2}$. None are detected by any competing pipeline except for CrossCorr in Ref. \cite{Messenger2015}. Viterbi tracking detects seven more signals (21, 50, 52, 54, 58, 71, 98) than TwoSpect, with $4.7 \times 10^{-27}$\,Hz$^{-1/2} \leq h_0/\sqrt{f_\star} \leq 1.46 \times 10^{-26}$\,Hz$^{-1/2}$ for these seven.

We have also verified that injections 41, 48, 57, 63, 64, 69, 72, 73 and 90 are not detected by the HMM, even when the nonwandering character of
the MDC Stage I signals is recognized explicitly by choosing a diagonal transition matrix $A_{q_iq_j} = \delta_{q_iq_j}$. This confirms that the Viterbi algorithm finds wandering and nonwandering signals with approximately equal efficiency, as long as the number of possible transitions at each step is relatively small (three or less here).

\begin{table*}
	\centering
	\setlength{\tabcolsep}{6pt}
	\begin{tabular}{lllllll}
		\hline
		\hline
		& CrossCorr & Viterbi & TwoSpect & Radiometer & Sideband & Polynomial\\
		\hline
		Hit rate (out of 50) & 50 & 41& 34 & 28 & 16 & 7\\
		Best $h_0$ ($10^{-25}$)  & 0.684 & 1.093& 1.250 & 2.237& 3.565& 7.678\\
		Best $h_0/\sqrt{f_\star}$ ($\times 10^{-25}$\,Hz$^{-1/2}$)& 0.020 & 0.047 & 0.082&0.102&0.235&0.261\\
		Typical $\varepsilon_{f_\star}$ (Hz) & $10^{-5}$ & $10^{-7}-10^{-3}$& $10^{-4}$ &$10^{-1}$& $10^{-2}$&$10^{-2}$\\
		Typical $\varepsilon_{a_0}$ (s) & $10^{-4}$ & $10^{-6}-10^{-1}$ & $10^{-2}$&$-$&$-$&$-$\\
		Typical run time (CPU-hr) &  $10^6$ & $10^3$ & $10^5$ & $10^3$ & $10^3$ & $10^8$\\
		\hline
		\hline
	\end{tabular}
	\caption{Comparison of Viterbi tracker and competing algorithms in Stage I of the Scorpius X-1 Mock Data Challenge \cite{Messenger2015}.}
	\label{tab:MDC_compare}
\end{table*}

\section{Conclusion}
In this paper, we describe an HMM  method for tracking a continuous gravitational wave signal with wandering spin frequency, emitted by either an isolated neutron star or a neutron star in a binary orbit. The HMM assumes a simple, nearest-neighbour-bin transition matrix combined with emission probabilities given by standard maximum likelihood matched filters: $\mathcal{F}$-statistic for an isolated target, and its Bessel-weighted version for a binary. The HMM is solved recursively for the optimal frequency history $f_\star(t)$ using the Viterbi algorithm. It is shown that, for Gaussian noise at a level characteristic of Advanced LIGO and with total observation time $T_\text{obs}=1$\,yr, the algorithm successfully tracks signals with $h_0\gtrsim 2 \times 10^{-26}$ (isolated), and $h_0 \gtrsim 8 \times 10^{-26}$ (binary). 

When applied to Stage I of the Scorpius X-1 Mock Data Challenge, the Viterbi tracker successfully detects 41 out of 50 synthetic signals with $\varepsilon_{f_\star(0)} < 4.26\times10^{-3}$\,Hz. In comparison, the CrossCorr, TwoSpect, Radiometer, Sideband and Polynomial algorithms detected 50, 34, 28, 16 and 7 out of 50 signals respectively. Performance metrics are summarized in Table \ref{tab:MDC_compare}. The frequency estimation error achieved by the Viterbi algorithm ranges from $\sim 10^{-3} {\rm Hz}$ to $\sim 10^{-7} \, {\rm Hz}$ in the first and second stages ($T_{\rm obs}=10\,{\rm d}$ and $1\,{\rm yr}$, two interferometers) and from $\sim 10^{-5} \,{\rm Hz}$ to $\sim 10^{-7} \, {\rm Hz}$ in the third stage ($T_{\rm obs}=1\,{\rm yr}$, three interferometers). In comparison, the frequency estimation errors achieved by the CrossCorr, TwoSpect, Radiometer, Sideband and Polynomial algorithms are of order $10^{-5}$\,Hz, $10^{-4}$\,Hz, $10^{-1}$\,Hz, $10^{-2}$\,Hz and $10^{-2}$\,Hz respectively. The $a_0$ estimation error achieved by the Viterbi algorithm spans the range $10^{-6} \leq \varepsilon_{a_0} / (1\,{\rm s}) \leq 0.6$. In comparison, only CrossCorr and TwoSpect estimate $a_0$ in the tests contained in Ref. \cite{Messenger2015}, achieving $\varepsilon_{a_0} \sim 10^{-4} \,{\rm s}$ and $10^{-2}\,{\rm s}$ respectively.

One advantage of the Viterbi tracker with respect to its competitors in the Mock Data Challenge is computational speed. For a 1-Hz band and $T_{\rm obs}=1\,{\rm yr}$, it takes $\sim 1$ CPU-hr to create $\mathcal{F}$-statistic data for $N_T = 36$ 10-d segments. It then takes $\sim 0.3$\,CPU-hr to process the $\mathcal{F}$-statistic data with the Viterbi tracker. Tracking both $f_\star$ and $a_0$ takes slightly longer than tracking $f_\star$ only, depending on the number of $a_0$ bins. In contrast, the CrossCorr and TwoSpect algorithms, which detected 50 and 34 out of 50 synthetic signals respectively, require $\sim 10^6$ CPU-hr to complete a typical broadband (0.5-kHz) search \cite{Messenger2015}. The computational savings offered by the Viterbi tracker can be re-invested to extend the astrophysical goals of the search, e.g. by searching a larger parameter space for Scorpius X-1 or targeting other low mass X-ray binaries \cite{Watts2008}.

Stage I of the Mock Data Challenge did not involve spin wandering. Nonetheless Viterbi tracking works exactly the same way, whether $f_\star(t)$ wanders or not. The algorithm is blind to the exact form of the transition matrix, so there is every reason to expect that the minimum $h_0$ detectable in Sections \ref{sec:isolated_tests} and \ref{sec:binary_tests} should carry over to realistic observations, with a possible caveat concerning nongaussian noise. At this stage, only Viterbi tracking has been tested systematically on spin wandering signals, successfully tracking sources with $h_0 \gtrsim 8 \times 10^{-26}$ and $\varepsilon_{f_\star(t)} < 5 \times 10^{-7}$\,Hz for noise levels representative of Advanced LIGO [$S_n(2f_\star)^{1/2} = 4\times 10^{-24} \, {\rm Hz^{-1/2}}$]. CrossCorr and TwoSpect \cite{Meadors2016} are also expected to track spin wandering sources well, but systematic testing is still underway.

\section{Acknowledgements}
We would like to thank Paul Lasky, Chris Messenger, Keith Riles, Karl Wette, Letizia Sammut, John Whelan and the LIGO Scientific Collaboration Continuous Wave Working Group for detailed comments and informative discussions.The synthetic data for Stage I of the Sco X-1 MDC were prepared primarily by Chris Messenger with the assistance of members of the MDC team \cite{Messenger2015}. We thank Chris Messenger and Paul Lasky for their assistance in handling the MDC data. L. Sun is supported by an Australian Postgraduate Award. The research was supported by Australian Research Council (ARC) Discovery Project DP110103347. 

\begin{table*}[!tbh]
	\centering
	\setlength{\tabcolsep}{2pt}
	\begin{tabular}{llllllllll}
		\hline
		\hline
		Index &	$f_\star$ (Hz) &	$h_0(10^{-25})$ & $h_0/\sqrt{f_\star}$ & $a_0$ & Estimated & Error in  & Estimated  & Error in  & Detection \\
		 &	 &	& ($10^{-25}$\,Hz$^{-1/2}$) & (sec) &  $f_\star$(Hz) & $f_\star$(Hz) &  $a_0$ (sec) & $a_0$ (sec) & Stage\\
		\hline
		1& 54.498391348174& 4.160& 0.563524& 1.379519& 54.4983906268& 7.214E-07& 1.3795200& 1.000E-06& 1\\
		2& 64.411966012332& 4.044& 0.503887& 1.764606& 64.4119658577& 1.546E-07& 1.7625600& 2.046E-03& 2\\
		3& 73.795580913582& 3.565& 0.415019& 1.534599& 73.795580441& 4.726E-07& 1.5523200& 1.772E-02& 1\\
		5& 93.909518008164& 1.250& 0.129012& 1.520181& 93.909517941& 6.716E-08& 1.5350400& 1.486E-02& 2\\
		11& 154.916883586097& 3.089& 0.248212& 1.392286& 154.91689756& 1.397E-05& 1.3996800& 7.394E-03& 2\\
		14& 183.974917468730& 2.044& 0.150706& 1.509696& 183.974917235& 2.337E-07& 1.0800000& 4.297E-01& 2\\
		15& 191.580343388804& 11.764& 0.849907& 1.518142& 191.580298596& 4.479E-05& 1.5148800& 3.262E-03& 1\\
		17& 213.232194220000& 3.473& 0.237865& 1.310212& 213.23225231& 5.809E-05& 1.3118575& 1.646E-03& 2\\
		19& 233.432565653291& 6.031& 0.394707& 1.231232& 233.432550338& 1.532E-05& 1.2297600& 1.472E-03& 2\\
		20& 244.534697522529& 9.710& 0.620916& 1.284423& 244.534696746& 7.765E-07& 1.2844800& 5.700E-05& 1\\
		21& 254.415047846878& 1.815& 0.113797& 1.072190& 254.415047445& 4.019E-07& 1.0724669& 2.769E-04& 3\\
		23& 271.739907539784& 2.968& 0.180071& 1.442867& 271.739936327& 2.879E-05& 1.4428800& 1.300E-05& 2\\
		26& 300.590450155009& 1.419& 0.081855& 1.258695& 300.59044964& 5.150E-07& 1.0800000& 1.787E-01& 2\\
		29& 330.590357652653& 4.275& 0.235096& 1.330696& 330.590357047& 6.057E-07& 1.3305600& 1.360E-04& 2\\
		32& 362.990820993568& 10.038& 0.526853& 1.611093& 362.990070589& 7.504E-04& 1.5926400& 1.845E-02& 1\\
		35& 394.685589797695& 16.402& 0.825579& 1.313759& 394.685618617& 2.882E-05& 1.3132800& 4.790E-04& 1\\
		36& 402.721233789014& 3.864& 0.192559& 1.254840& 402.721233202& 5.870E-07& 1.2556800& 8.400E-04& 2\\
		41& 454.865249156175& 1.562& 0.073240& 1.465778& 454.844343743& 2.091E-02& 1.4661565& 3.785E-04& \\
		44& 483.519617972096& 2.237& 0.101736& 1.552208& 483.519690961& 7.299E-05& 1.4601600& 9.205E-02& 2\\
		47& 514.568399601819& 4.883& 0.215277& 1.140205& 514.568399349& 2.528E-07& 1.1404800& 2.750E-04& 2\\
		48& 520.177348201609& 1.813& 0.079492& 1.336686& 520.327843196& 1.505E-01& 1.3370312& 3.452E-04& \\
		50& 542.952477491471& 1.093& 0.046897& 1.119149& 542.952491933& 1.444E-05& 1.1194380& 2.890E-04& 3\\
		51& 552.120598886904& 9.146& 0.389254& 1.327828& 552.120598435& 4.519E-07& 1.1431103& 1.847E-01& 2\\
		52& 560.755048768919& 2.786& 0.117639& 1.792140& 560.755063137& 1.437E-05& 1.7926028& 4.628E-04& 3\\
		54& 593.663030872532& 1.518& 0.062283& 1.612757& 593.663030722& 1.505E-07& 1.6131735& 4.165E-04& 3\\
		57& 622.605388362863& 1.577& 0.063198& 1.513291& 622.56610884& 3.928E-02& 1.5136818& 3.908E-04& \\
		58& 641.491604906276& 3.416& 0.134884& 1.584428& 641.491619251& 1.434E-05& 1.5848371& 4.091E-04& 3\\
		59& 650.344230698489& 8.835& 0.346437& 1.677112& 650.344215312& 1.539E-05& 1.6761600& 9.520E-04& 1\\
		60& 664.611446618250& 2.961& 0.114843& 1.582620& 664.611402246& 4.437E-05& 1.5840000& 1.380E-03& 2\\
		61& 674.711567789201& 6.064& 0.233463& 1.499368& 674.711611744& 4.395E-05& 1.5004800& 1.112E-03& 2\\
		62& 683.436210983289& 10.737& 0.410728& 1.269511& 683.436269138& 5.815E-05& 1.2700800& 5.690E-04& 1\\
		63& 690.534687981171& 1.119& 0.042584& 1.518244& 690.154763901& 3.799E-01& 1.5186360& 3.920E-04& \\
		64& 700.866836291234& 1.600& 0.060419& 1.399926& 701.313419622& 4.466E-01& 1.4002875& 3.615E-04& \\
		65& 713.378001688688& 8.474& 0.317256& 1.145769& 713.373737884& 4.264E-03& 1.0800000& 6.577E-02& 1\\
		66& 731.006818153273& 9.312& 0.344417& 1.321791& 731.006832222& 1.407E-05& 1.3219200& 1.290E-04& 1\\
		67& 744.255707971300& 4.580& 0.167871& 1.677736& 744.254311362& 1.397E-03& 1.0803545& 5.974E-01& 2\\
		68& 754.435956775916& 3.696& 0.134556& 1.413891& 754.435956631& 1.449E-07& 1.0800917& 3.338E-01& 2\\
		69& 761.538797037770& 2.889& 0.104699& 1.626130& 761.720204916& 1.814E-01& 1.6265499& 4.199E-04& \\
		71& 804.231717847467& 2.923& 0.103056& 1.652034& 804.231732078& 1.423E-05& 1.6524606& 4.266E-04& 3\\
		72& 812.280741438401& 1.248& 0.043792& 1.196485& 812.838541152& 5.578E-01& 1.1967940& 3.090E-04& \\
		73& 824.988633484129& 2.444& 0.085089& 1.417154& 825.182391835& 1.938E-01& 1.4175199& 3.659E-04& \\
		75& 862.398935287248& 7.678& 0.261467& 1.567026& 862.398964120& 2.883E-05& 1.1982204& 3.688E-01& 1\\
		76& 882.747979842807& 3.260& 0.109728& 1.462487& 882.74799427& 1.443E-05& 1.0796966& 3.828E-01& 2\\
		79& 931.006000308958& 4.681& 0.153408& 1.491706& 931.006764506& 7.642E-04& 1.4832000& 8.506E-03& 2\\
		83& 1081.398956458276& 5.925& 0.180165& 1.198541& 1081.39898556& 2.910E-05& 1.1980800& 4.610E-04& 2\\
		84& 1100.906018344283& 11.609& 0.349877& 1.589716& 1100.9060324& 1.406E-05& 1.2086045& 3.811E-01& 1\\
		85& 1111.576831848269& 4.553& 0.136553& 1.344790& 1111.57684611& 1.426E-05& 1.0800000& 2.648E-01& 2\\
		90& 1193.191890630547& 0.684& 0.019802& 1.575127& 1193.82006025& 6.282E-01& 1.5755337& 4.067E-04& \\
		95& 1324.567365220908& 4.293& 0.117966& 1.591685& 1324.56727666& 8.856E-05& 1.5926400& 9.550E-04& 2\\
		98& 1372.042154535880& 5.404& 0.145894& 1.315096& 1372.04216902& 1.448E-05& 1.0799668& 2.351E-01& 2\\
		\hline
		\hline
	\end{tabular}
	\caption{Results of Viterbi tracking of the 50 closed signals in the Sco X-1 Mock Data Challenge, sorted by index from Ref. \cite{Messenger2015}. The last column shows in which stage the injected signal is detected.}
	\label{tab:MDC}
\end{table*}

\end{document}